\documentclass[english,aps,nofootinbib,pra,notitlepage]{revtex4-1}
\pdfoutput=1
\usepackage[T1]{fontenc}
\usepackage[latin9]{inputenc}
\usepackage{color}
\usepackage{babel}
\usepackage{bbold}
\usepackage{dsfont}
\usepackage{amsmath}
\usepackage{graphicx}
\usepackage{subfig}
\usepackage{footnote}
\usepackage{amssymb}
\usepackage{float}
\usepackage{mathtools}
\usepackage[unicode=true, pdfusetitle,
 bookmarks=false,
 breaklinks=false,pdfborder={0 0 0},backref=false,colorlinks=false]
 {hyperref}
\usepackage{footmisc}
\makeatletter

\@ifundefined{textcolor}{}
{%
 \definecolor{BLACK}{gray}{0}
 \definecolor{WHITE}{gray}{1}
 \definecolor{RED}{rgb}{1,0,0}
 \definecolor{GREEN}{rgb}{0,1,0}
 \definecolor{BLUE}{rgb}{0,0,1}
 \definecolor{CYAN}{cmyk}{1,0,0,0}
 \definecolor{MAGENTA}{cmyk}{0,1,0,0}
 \definecolor{YELLOW}{cmyk}{0,0,1,0}
 }
\makeatother
\begin{document}

\preprint{Notes}
\title{Mirror-Field Entanglement in a Microscopic model for Quantum Optomechanics}
\preprint{Working draft: Not for distribution}

\author{Kanupriya Sinha}
\email{electronic address: kanu@umd.edu}
\affiliation{Joint Quantum Institute and Maryland Center for Fundamental Physics, University of Maryland, College Park, Maryland 20742, USA }
\author{Shih-Yuin Lin}
\affiliation{Department of Physics, National Changhua University of Education, Changhua 50007, Taiwan}
\affiliation{Department of Physics and Astronomy, University of Waterloo, Waterloo, Ontario, Canada N2L 3G1}
\author{B. L. Hu}
\affiliation{Joint Quantum Institute and Maryland Center for Fundamental Physics, University of Maryland, College Park, Maryland 20742, USA}

\begin{abstract}
%

We use a microscopic model, the Mirror-Oscillator-Field (MOF) model proposed in \cite{CR}, to describe the quantum entanglement between a mirror's center of mass (CoM) motion and a field. In contrast with the conventional approach where the mirror-field entanglement is understood as arising from the radiation pressure of an optical field inducing the motion of the mirror's CoM, the MOF model incorporates the dynamics of the internal degrees of freedom of the mirror that couple to the optical field directly. The major advantage in this approach is that it provides a self-consistent treatment of the three pertinent subsystems (the mirror's CoM motion, its internal degrees of freedom and the field) including their back-actions on each other, thereby giving a more accurate account of the quantum correlations between the individual subsystems. The optical and the mechanical properties of a mirror arising from its dynamical interaction with a quantum field are obtained without imposing any boundary conditions on the field additionally, as is done in the conventional way. As one of the new physical features that arise from this self-consistent treatment of the coupled optics and mechanics behavior we observe a coherent transfer of quantum correlations from the field to the mirror via its internal degrees of freedom. We find the quantum entanglement between the optical field and the mirror's center of mass motion upon coarse-graining over the internal degree of freedom. Further, we show that in certain parameter regimes the mirror-field entanglement is enhanced when the field interacts resonantly with the mirror's internal degree of freedom, a new result which highlights the importance of including the internal structure of the mirror in quantum optomechanical considerations.
\end{abstract}


\maketitle
\global\long\def\ket#1{\left|#1\right\rangle }
\global\long\def\bra#1{\left\langle #1\right|}
\global\long\def\bkt#1{\left(#1\right)}
\global\long\def\abs#1{\left\vert#1\right\vert}
\global\long\def\der#1#2{\frac{\mathrm{d}#1}{\mathrm{d}#2}}
\global\long\def\re{\mathrm{Re}}
\global\long\def\im{\mathrm{Im}}
\global\long\def\dd{\mathrm{d}}
\global\long\def\vect#1{\boldsymbol{#1}}
\global\long\def\avg#1{\left\langle #1 \right\rangle}
\global\long\def\leftexp#1#2{{\vphantom{#2}}^{#1}{#2}}
\global\long\def\mdf{\textit{mdf}}
\global\long\def\idf{\textit{idf}}
\global\long\def\th{$^{th}$}
\newcommand{\uv}[1]{\ensuremath{\mathbf{\tilde{#1}}}} 
\def\dblone{\hbox{$1\hskip -1.2pt\vrule depth 0pt height 1.6ex width 0.7pt
                  \vrule depth 0pt height 0.3pt width 0.12em$}}
\section {Introduction}
Optomechanics describes the interaction of light with mechanical systems. When an optical field interacts with a mechanical object there is a redistribution of the photon momentum upon reflection.  At the microscopic level this interaction results from the coupling of the electromagnetic (EM) field with the surface charges of the mechanical object (or the electrons in an atom). Still, the conventional approach towards studying optomechanical interactions only considers the effective boundary conditions for the optical field at the position of the mirror's CoM that arise from the microscopic picture in the steady state limit.  While the role of these internal degrees of freedom is universally acknowledged in the case of atom-field interactions when describing the mechanical effects of a field on an atom \cite{GordAsh, DaliCT}, their relevance in determining the optomechanical properties of larger systems is seldom discussed. 

However, questions pertaining to the transfer of quantum correlations,  such as calculating the quantum entanglement between the mirror and the field \cite{Asp13}--\cite{YC} or the superposition of two mirrors \cite{Marshall03},  demand that we take into consideration the full quantum nature of the macroscopic object including the dynamics of its quantal internal degrees of freedom (similar to the two-level atom) to systematically account for all the quantum correlations present between the individual subsystems. Such a treatment becomes one of a practical necessity when studying the optomechanical entanglement for well-isolated systems that preserve coherences for longer time scales, for example when considering the quantum entanglement between the motion of atoms or atomic ensembles and a field.

In a recent paper Galley, Behunin and one of the present authors \cite{CR} constructed a microphysics model called the mirror-oscillator-field (MOF) model that takes into consideration the microscopics of optomechanical interactions, providing a physically more complete theory for quantum optomechanics (QOM). The optical properties of the mirror are  captured  in this model  by introducing an internal degree of freedom of the mirror,  referred to as the mirror-oscillator, or \textit{mirosc} in \cite{CR}. The mirror as an \textit{optomechanical} element is described by two separate degrees of freedom corresponding to its center of mass motion (mechanics) and the surface charge that couples with the field (optics). We henceforth refer to these two degrees of freedom as the mechanical degree of freedom (\textit{mdf}) and the internal degree of freedom (\textit{idf}) respectively. The \textit{idf} and the \textit{mdf} are each depicted by a quantum oscillator, with the \textit{idf} coupled to an optical field that is modeled in \cite{CR} by a massless scalar field.  The \textit{idf} is what provides the indirect interaction between the field and the mirror's CoM motion, with its amplitude taking on field values at the position of the CoM.  Compared to the traditional approach where the effect of the mirror on the field is represented by imposing boundary conditions on the field at the position of the mirror, a microscopic treatment captures the mirror-field interaction in a more physically consistent way as both the internal and mechanical degrees of freedom of the mirror enter in determining the  dynamics self-consistently. As shown in \cite{CR}, different parameters of the \textit{idf} can describe a range of  optical activities, from broadband to narrow band reflectivity. With specific  parameter choices the authors in \cite{CR} made connections  to well-known optomechanical models including those of Barton \& Calogeracos \cite{BC}, Law \cite{Law} and Golestanian \& Kardar \cite{Kardar}.

The advantages of the MOF model over the usual practice of  imposing boundary conditions and the role of the \textit{idf} in capturing additional physical phenomenon is further expounded in this paper.  We also use the MOF model to study the quantum  entanglement between the mechanical motion of the mirror and the field. The conventional mechanism of mirror-field entanglement is by means of the radiation pressure coupling due to the photons of an optical field impinging on a mirror, transferring momentum to its center of mass \cite{Asp13}.  There is virtually no consideration of how the mirror's internal dynamics that give rise to its optical properties affect the entanglement of its external or mechanical degree of freedom with the field.  Even though the full description of this interaction at the microscopic level is quite complex, to gain a qualitative understanding of the coupled interplay of the optical and mechanical degrees of freedom the present relatively simple  MOF model  can serve the purpose aptly and economically.  As we shall see in this work, in some parameter regimes  the dynamics of a mirror's \textit{idf} play a nontrivial and even a decisive role in determining the transfer of correlations and hence the entanglement between the \textit{mdf} and the quantum field. We expect likewise for other related QOM effects.  

In the rest of this introduction we give a summary of the MOF model, followed by a description of the classical mechanical and optical properties of the MOF model  in Sec. II. Sec. III treats the quantum dynamics of the three interlinked subsystems -- the \textit{idf}, the \textit{mdf} and the field -- which leads to all the interesting physical phenomena in QOM. In particular we show that the usual radiation pressure coupling is  recovered as an approximation of the MOF model but one can go beyond these approximations to see new physical effects. With the solutions of the dynamical equations for this system we proceed in Sec. IV to derive the covariance matrix and calculate the entanglement between the mirror's center of mass and the quantum field. The role played by the internal degree of freedom  of the mirror is highlighted.  We conclude in Sec. V with a discussion of the main points.

\subsection*{The Mirror-Oscillator-Field (MOF) Model}
An optomechanical system consists of at least two components:  a mirror interacting with a quantum field -- where the "mechanics" refers to the mirror motion and "opto" refers to the field. One can think of three levels of description for this interaction: classical, semiclassical and quantum (see, e.g., \cite{RyanThesis}) in analogy to the studies of atom-field interaction. In the simplest classical electromagnetic description, the mirror is coupled to the field via the radiation pressure which can be obtained from the momentum flux imparted by the EM field that goes as the time-averaged Poynting vector of the field divided by the speed of light. In the semiclassical picture this force arises from the momentum transfer by the photons hitting the mirror, equivalently understood in terms of the gradient of energy density of the EM field on displacing the mirror. This leads to an interaction Hamiltonian of the intensity-position coupling form that goes as $\sim\hat N\hat x$, where $\hat N$ denotes the photon number operator and $\hat x$ the displacement of the mirror. In a microscopic description of the radiation pressure force, one can think of the field discontinuity at the mirror's CoM position inducing surface charge currents, which in turn experience a Lorentz force in the presence of the field that amounts to the radiation pressure force on the mirror's CoM \cite{ajp}. When one wishes to probe into issues like entanglement between a mirror and a field, a similar detailed treatment of the mirror and its interaction with the field is necessary, one that accounts for the transfer of quantum correlations between the mirror motion and the field as accurately as possible.

Consider, for example, a single atom as an optomechanical element whose\textit{ idf }is represented by a two level system. We know that the interaction between the field and the atom's two-level internal degree of freedom via photon emission and absorption is much stronger than the effective interaction of the field with the atom's center of mass degree of freedom. The coupling between the optical field and atomic motion  arises as the CoM motion alters the field configuration, thereby affecting the atom's internal level activities. Thus when dealing with the case of atoms as an optomechanical element \cite{SK12}--\cite{Meystre12}, one needs to regard its internal level dynamics with careful consideration. Similarly, it has also been shown that the atom's motional degree of freedom can affect the activities of its internal degrees of freedom such as spontaneous emission or motional decoherence as in \cite{Sanjiv03, Sanjiv032}. For this reason,  we point out the inadequacy in the study of mirror-field entanglement as one needs to take into consideration the \textit{idf} of the mirror that is essential in mediating the quantum correlations between the optical field and the mechanical motion. This was one of the primary motivations in the construction of the MOF model in \cite{CR}, which is highlighted in this work.

\begin{figure}[t]
\includegraphics[width = 4 in]{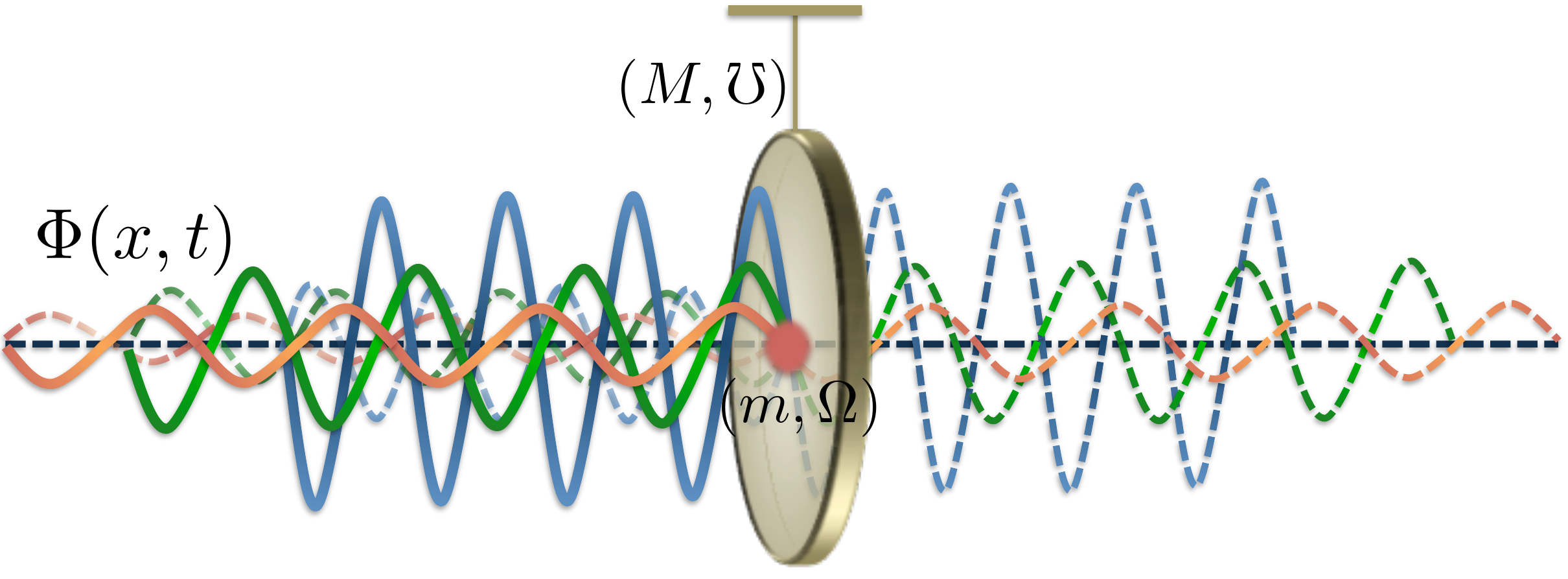}
\caption{Schematic representation of the interaction of a mirror with a field via its internal degree of freedom}
\label{schematic}
\end{figure}

Let us consider a point mirror interacting with a massless scalar field in (1+1)-dimensional space-time, the mirror is described by the two independent degrees of freedom - the \textit{mdf}  that has a mass $M$ and is suspended in a harmonic potential of frequency $\mho$ in addition to the \textit{idf} described by another harmonic oscillator of mass $m$ and frequency $\Omega$, as shown in Fig.\ref{schematic}.

While the \textit{mdf} does not interact with the field itself, the \textit{idf} is bilinearly coupled to the quantum field and constrained to be at the center of mass position leading to an effective interaction between the field and the \textit{mdf}, what we observe as the radiation pressure. The \textit{idf}-field interaction determines all the optical properties of the mirror as has also been studied in \cite{CR}. We assume that the \textit{idf}-field dynamics that represent the electronic excitations for the case of an atom happen at much faster time scales compared to those  of the mechanical motion of the atomic center of mass, such that $\Omega\gg\mho$.

For a non-relativistically moving mirror in the MOF model, the action is given by
\begin{align}
S = \int \dd t \bkt{\bkt{\frac{1}{2}M\dot{Z}^2-\frac{1}{2}M\mho^2Z^2}+\bkt{\frac{1}{2}m\dot{q}^2-\frac{1}{2}m\Omega^2q^2}+\int\dd x \frac{\epsilon_0}{2}\bkt{\bkt{\partial_t\Phi^2}-c^2\bkt{\partial_x\Phi}^2+\lambda\dot{q}\Phi\delta(x-Z)}}
\label{action}
\end{align}
where we denote the center of mass position  of the \textit{mdf} by $Z(t)$, the amplitude of the \textit{idf} by $q(t)$ and the scalar field by  $\Phi(x,t)$. For relativistic motion which is required for the treatment of acceleration radiation such as the Unruh effect, one needs to use the proper time,  modify the kinetic terms, and take care of the time-slicing scheme. Then the model will become a generalization of the Unruh-DeWitt detector theory \cite{RHA, RSG}. In drawing a correspondence between the scalar field and an electromagnetic field, we observe that the free field Lagrangian would correspond to that of an EM field if we choose $\Phi(x,t)$ to represent the vector potential $A$. We have chosen a form of the bilinear coupling motivated by the electrodynamic form of interaction $\bkt{\sim \frac{e}{mc} p \cdot A}$, bearing in mind that the mirror's \textit{idf} can potentially represent the electronic level structure inside an atom. We note that this is different from the form of coupling in the original MOF model \cite{CR} $\bkt{\sim\lambda q\Phi}$.  The $\delta(x-Z)$ factor in the coupling restricts the \textit{idf}-field interaction to the center of mass position and the position dependence of the scalar field in turn leads to an effective force on the \textit{mdf}. We choose the coupling $\lambda$ to have the dimensions of the electronic charge $e$ and $\Phi(x,t)$ to have the dimensions of $A/c$. This is in agreement with the correspondence of the MOF model with the Barton-Calogeracos (BC) model \cite{BC}, where in the limit of adiabatic \textit{idf} evolution the coupling $\lambda$ can be physically identified as the surface charge density. Noticing that the free space permittivity in (1+1)-dimensions scales as $\epsilon_0\sim (\text{Charge})^2 \text{(Time)}^2\text{(Mass)}^{-1}\text{(Length)}^{-1}$, the free field Lagrangian in \cite{CR} has been rescaled here by a factor of $\epsilon_0$ for dimensional consistency.

\section{Classical Optomechanical properties}

In this section we will illustrate how the MOF model can describe the classical optical and mechanical properties exhibited by a mirror, leading to the known intensity-position radiation pressure coupling. We begin with deriving the coupled equations of motion for the classical amplitudes of the \textit{mdf}, \textit{idf} and field ($\{\bar{Z},\dot{\bar{Z}},\bar{q},\dot{\bar{q}},\bar{\Phi},\dot{\bar{\Phi}}\}$ respectively) from the action in \eqref{action} ($\delta S =0$)

\begin{align}
&\ddot{\bar{Z}}+\mho^2\bar{Z} = \frac{\lambda\dot{\bar{q}}}{M}\partial_x\bar{\Phi}\bkt{\bar{Z},t}\label{class1}\\
&\ddot{\bar{q}}+\Omega^2\bar{q} = -\frac{\lambda}{m}\dot{\bar{\Phi}}\bkt{\bar{Z},t}\label{class2}\\
&\epsilon_0\bkt{\ddot{\bar{\Phi}}\bkt{x,t}-c^2\partial_x^2\bar{\Phi}\bkt{x,t}} = \lambda\dot{\bar{q}}\delta(x-\bar{Z})\label{class3}
\end{align}

It can be seen that the moving \textit{idf} acts as a point source for the field and the \textit{idf} is in turn driven by $\dot\Phi$ at the center of mass position $\bar{Z}$, which in the  electromagnetic correspondence represents the electric field at the CoM position $\bkt{\dot{\Phi}\sim E}$. Also, with $\lambda$ representing the charge density, it can be seen from \eqref{class2} that the force on the surface charge degree of freedom goes as $\sim \lambda \dot{\bar\Phi}$.  We have assumed here that the mirror center of mass velocity is in the non-relativistic limit, such that $\left\vert\frac{\dd \bar{Z}}{\dd t}\right\vert\ll c$. For a relativistically moving mirror, the \textit{idf} would more generally observe a Doppler shift of the field with respect to the moving center of mass as \eqref{class2} becomes
\begin{align}
\ddot{\bar{q}}+\Omega^2\bar{q} = -\frac{\lambda}{mc} \left( \dot{\bar{Z}}^0  \partial_t 
  + \dot{\bar{Z}}^1 \partial_x \right)\bar{\Phi}(\bar{Z}^\mu) 
\end{align}
where $\bar{Z}^\mu = (\bar{Z}^0(\tau), \bar{Z}^1(\tau))$ is the worldline of the mirror parametrized by its proper time $\tau$, and $\dot{O} \equiv dO/d\tau$. As the motion of the mirror center of mass leads to the motion of the charges sitting on the surface that interact with the field, the surface charges experience a Doppler shifted field which in turn changes their optical response leading to dynamically changing boundary conditions observed by the field.  

If one prefers to think in terms of applying boundary conditions on the field, in the MOF model it would correspond to the steady state response of the internal degree of freedom. Thus our model captures the full dynamical interplay as opposed to the static condition in the conventional approach. In fact, a simple generalization of the set up here can deal with a relativistically moving mirror as in dynamical Casimir effect (DCE), whereas the conventional method of imposing a static boundary condition on the field would fail to address dynamical situations wherein the time scales of the mechanical motion are comparable those of {\it idf}-field interaction dynamics. \footnote{This is an important point long explored and resolved in cosmological particle creation which results from the same mechanism but with the expanding universe playing the role of an external agent as in DCE.}

For now, we restrict our attention to a non-relativistically moving mirror.  For the case where the system dynamics is driven by an incident field (and not by any external agent which accelerates the mirror as is in the setup of the Unruh effect), this is ensured from the separation of the timescales for the internal and center of mass degrees of freedom $\bkt{\Omega\gg\mho}$. We will demonstrate this further in  Section \ref{OP} for the case of a single mode field.

Knowing the coupled system dynamics, below we first look at  how the radiation pressure force arises from our model in the non-relativistic limit.

\subsection{Classical Radiation Pressure Force}
\label{RP}
As seen from \eqref{class1}, the \textit{mdf} is driven non-linearly by both the \textit{idf} and the field. We now eliminate the \textit{idf} from the picture to obtain the mechanical force on the center of mass.

From integrating the field equation of motion \eqref{class3} around the mirror center of mass position $\bar Z$, we see that there is a discontinuity in the field spatial derivative. This can be understood as the discontinuity in the magnetic field across the mirror surface in the electromagnetic correspondence ($\partial_x\Phi\sim B$) coming from the surface charge current $\dot{\bar{q}}$.
In the non-relativistic limit we find the surface charge current as
\begin{align}
\lambda\dot{\bar{q}} = -\epsilon_0c^2\partial_x\bar{\Phi}\vert_{\bar{Z}^-}^{\bar{Z}^+}
\label{classsc}
\end{align}
The surface charge current being induced by the discontinuous magnetic field across the center of mass position can be interpreted as the Ampere's law in 1+1 dimensions. We eliminate the \textit{idf} from the center of mass dynamics, defining the spatial derivative of the field at the center of mass position as  $\partial_x\bar \Phi\bkt{\bar Z,t}\equiv\bkt{\partial_x\bar\Phi\bkt{\bar Z^+,t}+\partial_x\bar\Phi\bkt{\bar Z^-,t}}/2$. We rewrite CoM dynamics  as 
\begin{align}
\ddot{\bar{Z}}+\mho^2\bar{Z} = -\frac{1}{2M}\epsilon_0c^2(\partial_x\bar{\Phi})^2\vert_{\bar{Z}^-}^{\bar{Z}^+}
\label{rp}
\end{align}
We notice that the right hand side corresponds to the well-known radiation pressure force $\bkt{\sim \frac{B^2}{2\mu_0}}$ seen by a mirror in the non-relativistic limit \cite{Law}. This is justified based on the fact that the electric field vanishes at the mirror position in the co-moving reference frame and the force being proportional to the EM field energy density then goes as $\sim \frac{B^2}{2\mu_0}$. To compare with the expression in \cite{Law},  we notice that for a perfect mirror there is no field energy density on one side of the mirror $\bkt{\partial_x\Phi\bkt{\bar Z^+,t}=0}$ and we reduce to the known result. For an imperfect mirror there is a finite energy density of the EM field on either side of the surface, hence the net radiation pressure force is given by the difference in the field energy density on either side of the surface as in \eqref{rp}. In the MOF model the radiation pressure force can be interpreted as the Lorentz force arising from the interaction of the induced surface charge current \eqref{classsc} with the magnetic field $(\partial_x\Phi)$. Such an interpretation of the radiation pressure force as the Lorentz force on induced surface charge currents has been discussed in detail in \cite{ajp}.

Thus we have arrived at the classical radiation pressure force on the mirror in the non-relativistic CoM motion limit as one would find from imposing fixed boundary conditions on the field. Rather, in this case the boundary conditions resulting from the mirror-field coupling arise self consistently from the dynamical interaction between the moving \textit{idf} and the field, as does the radiation pressure. We note here that while the form of the radiation pressure force we obtain from including the \textit{idf} is identical to what we get from imposing the fixed boundary conditions, the boundary conditions themselves rather than being fixed are determined by the dynamics of the \textit{idf}-field interaction. This more generally includes retarded influence of the moving surface charges on the field in a dynamical way. To see this more concretely, consider the \textit{idf} amplitude solution from \eqref{class2}
\begin{align}
\bar{q}(t) = \bar{q}_h(t)+\int_0^t\dd t' G_i(t-t') \bkt{-\frac{\lambda}{m}\dot{\bar{\Phi}}\bkt{\bar{Z}(t'),t'}}
\label{idfsol}
\end{align}
where we define $\bar{q}_h$ as the homogeneous solution for the free \textit{idf} evolution and $G_i(t-t') \equiv \frac{\sin(\Omega\bkt{t-t'})}{\Omega}$ as the Green's function for the \textit{idf}. We use this to eliminate the \textit{idf} from the field's equation of motion to get
\begin{align}
\epsilon_0\bkt{\partial_t^2-c^2\partial_x^2}\bar{\Phi}\bkt{x,t} +\frac{\lambda^2}{m}\delta\bkt{x-\bar{Z}(t)}\int_0^t\dd t'\partial_tG_i\bkt{t-t'}\dot{\bar{\Phi}}(\bar{Z}(t'),t')= \lambda\dot{\bar{q}}_h\delta\bkt{x-\bar{Z}(t)}
\end{align}
We see that the \textit{idf} is driven by the field and influences the field in return, as captured in the second term on the left hand side that represents the retarded influence of the \textit{idf} on the field, meaning that the radiation pressure force depends on the coupled non-Markovian dynamics of the field, center of mass and the \textit{idf}. Thus, we can identify the term $\partial_tG_i\bkt{t-t'}\equiv\chi\bkt{t-t'}$ as the susceptibility function for the mirror. To compare with the case where one applies boundary conditions as opposed to including the \textit{idf} dynamics self-consistently one needs to include the coupling of the \textit{idf} with a bath so as to reach the steady state response of the damped \textit{idf}. We will further illustrate this point and the role of the internal degree of freedom in determining the optical properties of the mirror in the following subsection.

\subsection{Optical properties}
\label{OP}
To study the optical properties arising from the MOF model let us consider a single mode field at frequency $\omega$ and amplitude $\Phi_0$ driving the mirror's \textit{idf}. Assuming that the mirror's CoM is at the origin in equilibrium, we make the following plane-wave ansatz for the field
\begin{align}
\Phi_\omega\bkt{x,t} = \frac{\Omega}{\omega}\Phi_0e^{-i\omega t}\bkt{\Theta(-x)\bkt{e^{ikx}+R(\omega)e^{-ikx}}+\Theta(x)T(\omega)e^{ikx}}+H.C.
\label{pwa}
\end{align}
where we have introduced the frequency normalization factor $\bkt{\Omega/\omega}$ to take care of the fact that in the EM correspondence the electric field amplitude $\bkt{E \sim\partial_tA}$ is independent of the frequency of the field. $R\bkt{\omega}$ and $T\bkt{\omega}$ refer to the reflection and transmission coefficients of the point mirror, such that $T\bkt{\omega} = 1+R\bkt{\omega}$. In considering the interaction of the {\it idf} with only a single field mode, we include a damping ($\gamma_f$) that arises from its coupling with the remaining field modes. For current purposes, we assume that the damping is small $\bkt{\gamma_f\ll\Omega}$ so that one can ignore the dissipation of the incident plane wave. As in \cite{CR}, we assume that in the steady state regime the \textit{idf} oscillates at the frequency of the incident field. In which case, we find
\begin{align}
q(t) = \frac{-i\omega\lambda}{m\bkt{\omega^2-\Omega^2}}\frac{\Omega}{\omega}\Phi_0T(\omega)e^{-i\omega t}+H.C.
\end{align}
From the mirror center of mass dynamics \eqref{rp}, we can see that in the presence of the incident drive the center of mass consists of a time-independent and a high frequency ($2\omega$) radiation pressure term, coming from the non-linear interaction of the incident B field and the induced surface charge current.  In the limit $\mho\ll\Omega$, the high frequency component of the mirror amplitude denoted by $\bar{Z}_{2\omega}$ scales as $|\bar{Z}_{2\omega}|\sim\frac{\epsilon_0\Phi_0^2 \Omega^2}{M \omega^2}$, which, in the near field-{\it idf} resonance regime $(\omega\approx\Omega)$, is much smaller compared with the mirror amplitude coming from the constant radiation pressure part $\bar{Z}_0\sim\frac{\epsilon_0\Phi_0^2\Omega^2}{M\mho^2}$, noting that $\frac{\bar Z_0}{\bar Z_{2\omega}}\sim\frac{\Omega^2}{\mho^2}\gg1$. Thus we find that the mirror position evolves essentially at its natural frequency $\mho$ under the constant force. 

We assume that at the classical level the center of mass motion does not affect the \textit{idf}-field coupling and the resulting optical properties from the interaction. More explicitly, the phase of the field mode that is resonant with the \textit{idf} changes by a very small amount over the length scales of one amplitude of the \textit{mdf}, that is $\Delta \phi \equiv \bkt{\Omega/c}\bar{Z}_0\ll1$. This restricts the field amplitude to 
\begin{align}
\abs{\Phi_0}^2\ll\frac{M\mho^2c}{\Omega^3\epsilon_0}
\label{SAA}
\end{align}
This is a self-consistent validity constraint which ensures that the optical properties of the mirror are unaffected by the center of mass motion to first order, to reaffirm our plane wave ansatz \eqref{pwa}. Physically speaking we assert that the mirror CoM motion is much smaller than the wavelengths of the field that it interacts with. The sub-wavelength motion approximation is valid for the case of trapped atoms spatially confined in a harmonic trap (trap frequency being $\mho$ in this case), interacting with an optical field of frequency $\omega$.

In the plane wave ansatz, we find the surface charge current for the \textit{idf} \eqref{classsc} as 
\begin{align}
\lambda\dot{\bar{q}} = -\epsilon_0c^2\partial_x\bar{\Phi}\bkt{x,t}\vert_{\bar{Z}^-}^{\bar{Z}^+} &\approx -2ik\epsilon_0c^2\Phi_0\frac{\Omega}{\omega}e^{-i\omega t}R\bkt{\omega}+H.C.\nonumber\\
& = -2i\epsilon_0\Omega c\Phi_0e^{-i\omega t}R\bkt{\omega}+H.C.
\label{sfc}
\end{align}
We can notice here that the induced surface charge current is proportional to the mirror reflectivity. Thus, as expected, a higher reflectivity leads to a larger radiation pressure force.

Now within the non-relativistic and sub-wavelength CoM motion approximations, we consider the MOF model with the two different forms for the coupling term - (1) $q\Phi$ (as previously analyzed in \cite{CR}) and (2) $\dot{q}\Phi$ - and study the optical properties that arise from these two couplings in different parameter regimes.

\subsubsection{$q\Phi$ coupling}
Let us first consider the $q\Phi$ coupling as in \cite{CR} and start with drawing the correspondence between the interaction term for the scalar field vis-a-vis an EM field. As motivated in the section II.B.1 in \cite{CR} when comparing the MOF model with the Barton-Calogeracos (BC) model, we choose the coupling $\lambda$ to have dimensions of the charge density such that dimensionally $\lambda\sim \text{(Charge)}\text{ (Length})^{-1}$. Going back to the interaction term in the original action we use this to find the dimensions of the scalar field as $\Phi\sim (\text{Mass}) \text{(Length)}^2\text{(Time)}^{-2}/\text{(Charge)}$ and rescale the free field term accordingly, we get for the free field action
\begin{align}
S_F = \frac{\epsilon_0}{2c^2}\int\dd t\int\dd x \bkt{\bkt{\partial_t\Phi}^2-c^2\bkt{\partial_x\Phi}^2}
\end{align}
This leads to the coupled {\it idf}-field equations of motion for a fixed center of mass as
\begin{align}
\label{qp1}
\epsilon_0/c^2\bkt{\partial_t^2\Phi-c^2\partial_x^2\Phi }&= \lambda q\delta(x)\\
\label{qp2}
m\ddot{q}+m\Omega^2q &= \lambda\Phi(0,t)
\end{align}

\begin{figure}[ht]
\subfloat[]{\includegraphics[width=3.5 in]{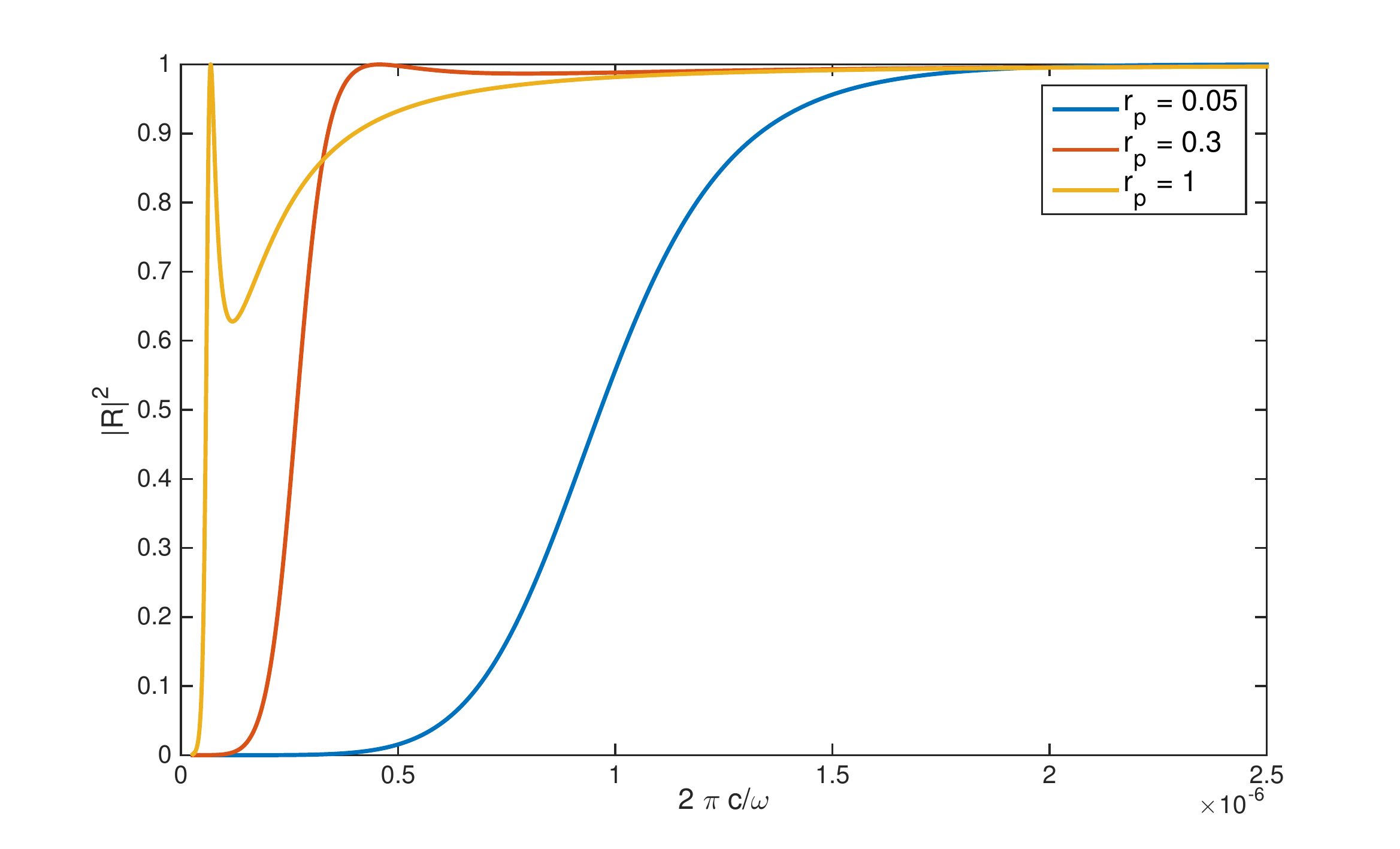}}
\subfloat[]{\includegraphics[width=3.5 in]{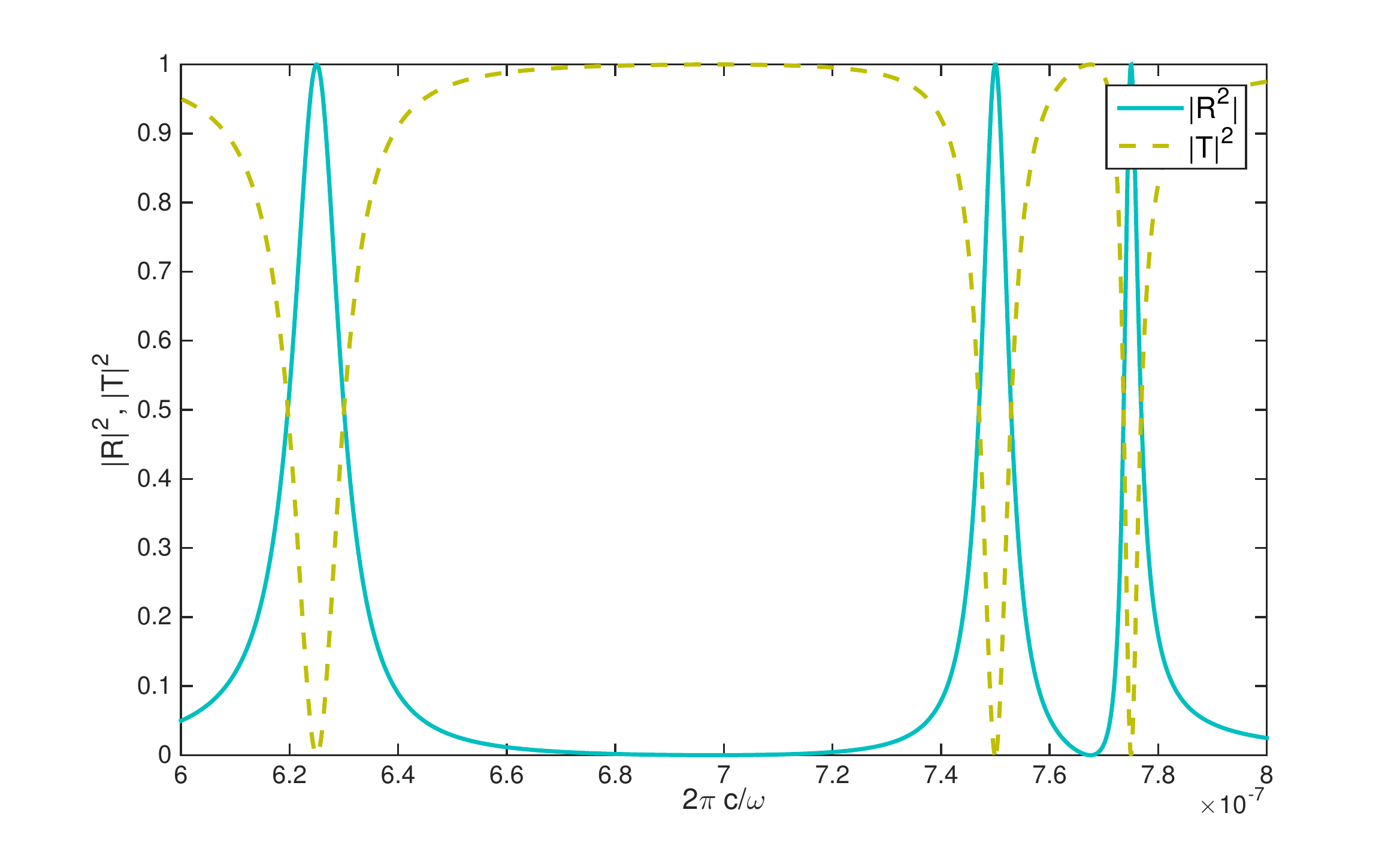}}
\caption{Reflection properties from the two different forms of coupling ($q\Phi$ and $\dot q \Phi$) (a) Reflectance as a function the incident field wavelength for different \textit{idf} to plasma frequency ratios $(r_p = \Omega/\Omega_P)$, the plasma frequency is fixed at $\Omega_P = 1.37\times10^{16}$  Hz (for silver) from $q\Phi$ coupling, choosing $r_p \approx 0.3$ mimics the cut-off behavior for silver (b) Reflectance and transmittance spectrum from $\dot q \Phi$ coupling to simulate the optical response for a photonic crystal as from the experimental results in \cite{phc}. Each resonance corresponds to a separate {\it idf}.}
\label{refspec}
\end{figure}

For a plane wave incident on the mirror, using the ansatz \eqref{pwa} to solve for the surface charge current as in \eqref{sfc} in the steady state limit we get the reflectivity for the case of $q\Phi$ coupling as
\begin{align}
R(\omega) = \frac{-i\lambda^2c}{i\lambda^2c+2m\omega\epsilon_0(\omega^2-\Omega^2)};\quad\abs{R}^2 =  \frac{1}{1+r_p^2 \eta^2\bkt{1-\eta^2}^2}
\end{align}
where we have defined the ratio of the field to the {\it idf} frequency as $\eta\equiv\omega/\Omega$ and $r_p\equiv\bkt{\frac{2m\Omega^3\epsilon_0}{\lambda^2c}}\equiv\Omega/\Omega_P$. We identify the quantity $\Omega_P\equiv\frac{\lambda^2c}{2m\Omega^2\epsilon_0}$ as the plasma frequency, again motivated by the comparison with the BC model. As found in \cite{CR}, the mirror becomes perfectly reflecting for 1) infinitely strong \textit{idf}-field coupling, $\lambda\rightarrow\infty$,  2) perfect resonance between the \textit{idf} and the incident field, $\omega=\Omega$ or 3) massless \textit{idf}, $m\rightarrow0$. Now observing that the reflection spectrum is completely characterized by the two frequency ratios $r_p$ (ratio of the \textit{idf} to plasma frequency) and $\eta$ (ratio of the field to \textit{idf} frequency), we consider different values for the parameter $r_p$ and look at the reflectance as a function of the field frequency for a fixed plasma frequency, as shown in Fig. \ref{refspec}.

To invoke the correspondence with the BC model \cite{BC} we need to assume that the \textit{idf} evolves adiabatically in the limit $\{m\rightarrow0,\Omega\rightarrow\infty\}$ such that the quantity $m\Omega^2\equiv\kappa$ that physically corresponds to the mass density of the surface charges stays finite. In this limit since $r_p\gg1$ ($\Omega\rightarrow\infty$), we see a resonant behavior in the reflection spectrum around the \textit{idf} frequency $\Omega$. In the regime where $r_p\ll1$, the reflection spectrum shows a high frequency cutoff behavior similar to the case of bulk metals with Drude-Lorentz response. As shown in Fig.\ref{refspec}, given the plasma frequency for silver ($\Omega_P = 1.37\times 10^{16}$ Hz), we compare the known optical response with our model and find that a \textit{idf} to plasma frequency ratio $r_p \approx 0.3$  mimics the cut-off behavior reasonably well. Knowing that the charge carrier density for silver is $n_s= 5.8 \times10^{28} m^{-3}$ and using the BC correspondence to find $\lambda = n_s e$, we can deduce all three \textit{idf} parameter values.

\subsubsection{$\dot{q}\Phi$ coupling}
As in the previous subsection we find the reflection coefficient for the $\dot q \Phi$ coupling as
\begin{align}
R(\omega) = \frac{-i\lambda^2\omega}{i\lambda^2\omega+2m\epsilon_0c(\omega^2-\Omega^2)};\quad \abs{R(\omega)}^2 =\frac{\omega^2/\Omega^2}{\omega^2/\Omega^2+\bkt{\frac{2m\epsilon_0c\Omega}{\lambda^2}}^2(\omega^2/\Omega^2-1)^2} = \frac{\eta^2}{\eta^2+r_p^2(\eta^2-1)^2}
\label{trans}
\end{align}
where again we have defined the ratio of the field to \textit{idf} frequency as $\eta\equiv\omega/\Omega$ and the ratio of \textit{idf} to plasma frequency as $r_p\equiv \Omega/\Omega_P=\Omega/(\frac{\lambda^2}{2m\epsilon_0c})$, redefining the plasma frequency as $\Omega_P=\frac{\lambda^2}{2m\epsilon_0c}$. We see that mirror becomes perfectly reflecting for the same conditions as in the case of $q\Phi$ coupling $\{\lambda\rightarrow\infty,\omega = \Omega, m\rightarrow 0\}$. Unlike the $q\Phi$ coupling, we do not see a perfect reflection at $\omega = 0$ which was an artifact of the monopole coupling between the \textit{idf} and the field. 

The optical response exhibits a resonant behavior around the \textit{idf} frequency $\Omega$, since the reflectivity is maximum for $\eta=1$. For this reason it is natural to consider optomechanical elements with built in resonances such as photonic crystals or atoms as an application. We find that one can mimic the optical response of a photonic crystal structure (see Fig.\ref{refspec}) by choosing the resonant frequency of the \textit{idf} as the resonant mode of the photonic crystal, for multiple resonances we choose multiple internal degrees of freedom such that $\Omega_i = \omega_i^{res}$ at each resonance peak. The sharpness of the resonance is determined by the quantity $r_p^i$, since the parameter $r_p^i$ determines the coupling strength of the field to a particular resonance mode of the structure. Thus one can determine the two parameters that characterize the optical response in our model, namely $r_p$ and $\Omega$. To completely determine all the parameters of the internal  degree of freedom $\{m,\Omega, \lambda\}$ we need to draw an additional physical correspondence between the internal degree of freedom and the physical setup as we did for the previous case of $q\Phi$ coupling by identifying the coupling constant $\lambda$ as the charge density.

As we had noticed previously, the mirror reflectivity characterized by the {\it idf} parameters determines the strength of the induced surface charge current \eqref{sfc} which in turn factors into determining the radiation pressure coupling. In the following section we will show that the same applies to the case of coupling between the quantum fluctuations of the mirror and the field. We now turn to look at the coupled quantum dynamics of the three subsystems in the MOF model.
%

\section{Quantum Dynamics of the Coupled Mirror-Oscillator-Field (MOF) System}

Let us perturb the original action \eqref{action} about the classical solutions as $\{\bar{Z}+\tilde{Z}, \bar{q}+\tilde{q},\bar{\Phi}+\tilde{\Phi}\}$, with $\tilde O$ being the deviations about the classical solutions $\bar O$. Assuming that the center of mass motion about $\bar{Z}$ is small and restricted to the sub-wavelength regimes $\bkt{k\tilde Z\ll1}$ for the field modes below a certain high frequency cutoff, we expand the action up to third order in the fluctuations about the classical solutions ignoring terms that are second order or higher in $k \tilde Z$. We go up to third order to specifically include the term that couples the perturbations of all three subsystems  (labeled as MOF below) to arrive at the non-linear intensity-position $\bkt{\sim\hat N \hat x}$ coupling. In the subsequent dynamics we shall only consider bilinear interaction terms to preserve Gaussianity of the individual subsystems. As we will see, truncating the action up to second order corresponds to the linearized approximation in the limit of strong mean-field amplitude, also called a background field expansion in field theory. 

We write the perturbed action as

\begin{align}
S_3 = &\int\dd t\underbrace{\bkt{ \frac{1}{2} M\dot{\tilde{Z}}^2-\frac{1}{2} M\mho^2\tilde{Z}^2}}_{\text{mdf (M)}}+\underbrace{\bkt{\frac{1}{2} m\dot{\tilde{q}}^2 - \frac{1}{2} m\Omega^2\tilde{q}^2}}_{\text{idf (O)}}+\int\dd x\left[\underbrace{\frac{\epsilon_0}{2}\bkt{\bkt{\partial_t\tilde \Phi}^2-c^2\bkt{\partial_x\tilde \Phi}^2 }}_{\text{Field (F)}}+\lambda \delta\bkt{x-\bar Z}\right.\nonumber\\
&\left.\bkt{\underbrace{\dot{\tilde q}\tilde \Phi}_{\text{OF}} +\underbrace{ \dot {\bar q}\bkt{\partial_x\tilde\Phi} \tilde Z }_{\text{MF}}+\underbrace{\dot{\tilde q}\bkt{\partial_x\bar \Phi }\tilde Z}_{\text{OM}}+\underbrace{\dot{\tilde q}\bkt{\partial_x\tilde \Phi}\tilde Z}_{\text{MOF}}}\right]
\label{s3}
\end{align}
One can observe several points from the above expression, firstly, we find that there is an effective coupling between the fluctuations of the mirror center of mass and the field via the internal degree of freedom as denoted by the terms MF and MOF.  To the lowest order, the mirror-field coupling strength is proportional to the classical surface current $\dot{\bar q}$, implying that the fluctuations of the field are the most sensitive to the fluctuations of the mirror center of mass if the surface current is at its largest.  In the single field mode case this is proportional to the reflection coefficient of the mirror as seen in \eqref{sfc}, meaning that a highly reflecting mirror leads to large effective MF coupling strength. Secondly, there is also an effective coupling between the \textit{idf} and the \textit{mdf} fluctuations denoted by the terms OM and MOF, which to the lowest order is proportional to the spatial derivative of the field (or B) at the center of mass position. The coupling strengths of the interaction terms between the {\it idf} and the mirror (OM), and the field and the mirror (MF) are determined by the classical solutions of the field and \textit{idf} amplitudes as found in the previous sections.

We get the following equations of motion for the coupled mirror and field dynamics
\begin{align}
&\ddot{\tilde Z}+\mho^2\tilde Z = \frac{\lambda}{M}\left[\dot{\bar q}\partial_x\tilde \Phi\bkt{\bar Z,t}+\dot{\tilde q}\left\{\partial_x\bar \Phi \bkt{\bar Z,t}+\partial_x\tilde \Phi \bkt{\bar Z,t}\right\}\right]\label{qM}\\
&\epsilon_0\bkt{\partial_t^2\tilde\Phi - c^2\partial_x^2\tilde \Phi} = \lambda \dot{\tilde q} \delta\bkt{x-\bar Z} -  \lambda\bkt{ \dot{\bar q} + \dot{\tilde q} }\partial_x \bkt{\delta\bkt{x-\bar Z} }\tilde{Z}\label{qF}
\end{align}
It can be seen here that unlike the classical equations of motion, the field fluctuations are not only driven by the {\it idf} but also by the fluctuations of the center of mass position. From integrating \eqref{qF} around the classical center of mass position $\bar Z$, we get the surface current fluctuation as
\begin{align}
\lambda\dot{\tilde q} = -\epsilon_0c^2 \partial_x\tilde \Phi \vert_{\bar Z_-}^{\bar Z_+}
\label{qsc}
\end{align}
just as the classical version interpreted as the Ampere's Law in 1+1 D in \eqref{classsc}. Using this and the classical surface current to eliminate the \textit{idf} from the center of mass dynamics \eqref{qM}, we get
\begin{align}
\ddot{\tilde Z}+\mho^2\tilde Z = \frac{-\epsilon_0c^2}{M}\left[ \left.\bkt{\partial_x\bar \Phi}\bkt{ \partial_x\tilde\Phi}\right\vert_{\bar Z_-}^{\bar Z_+}+\frac{1}{2}\left.\bkt{ \partial_x\tilde \Phi}^2 \right\vert_{\bar Z_-}^{\bar Z_+}\right]
\end{align}
We can see that the first term on the right side corresponds to the radiation pressure coupling in the linearized approximation which is valid for large photon numbers in the presence of a classically driven field. The second term goes beyond this approximation, which corresponds to the $\hat N\hat x$ type of coupling, required for treating situations with small photon numbers. Considering $\tilde \Phi$ represents the quantum fluctuations of the field, we can understand the radiation pressure force at the quantum level as arising from the asymmetry in the field fluctuations on either side of the mirror. Say, if there were a cavity present on one side and free space on the other, the radiation force from the cavity side would be stronger in comparison because of the small quantization volume leading to asymmetry in the density of field modes as in the case of Casimir force \cite{Casimir}. Such an interpretation of Casimir force as a radiation pressure force from the vacuum field has been discussed by Milonni {\it et al} in \cite{Milonni98} for the case of two perfectly conducting plates.

We now restrict ourselves to second order perturbations in the original action, to keep all the interaction terms bilinear such that starting out with Gaussian initial states for the three subsystems, Gaussianity of the individual subsystems is preserved. We derive the conjugate momenta from the second order action as
\begin{align}
\tilde p  &= m\dot{\tilde q}+\lambda\tilde \Phi\bkt{\bar Z,t} +\lambda \partial_x\bar\Phi(\bar Z,t)\tilde Z\\
\tilde P &= M\dot{\tilde Z}\\
\tilde\Pi (x,t) &= \epsilon_0\dot{\tilde\Phi} (x,t)
\end{align}

It can be seen that the fluctuations in the \textit{idf} are influenced by both the \textit{mdf} and the field and hence mediate the effective interactions between the two. Identifying the dynamical variables $\{\tilde Z,\tilde q,\tilde \Phi\}$ as the quantum fluctuations of the {\it mdf}, {\it idf} and the field respectively about their mean-field amplitudes, we arrive at the second order Hamiltonian 
\begin{align}
\tilde{H}_2\equiv&\underbrace{\frac{\tilde P^2}{2M} + \frac{1}{2} M\bkt{\mho^2+\frac{\lambda^2}{mM}\bkt{\partial_x\bar\Phi(\bar Z,t)}^2}\tilde Z^2}_{mdf (M)}+\underbrace{\frac{\tilde p^2}{2m}+\frac{1}{2}m\Omega^2\tilde q^2}_{\text{idf} (O)}\nonumber\\
&+\underbrace{ \int\dd x \bkt{\frac{\tilde \Pi^2}{2\epsilon_0}+\frac{1}{2}\epsilon_0c^2\bkt{\partial_x\tilde\Phi}^2}+\frac{\lambda^2}{2m}\tilde\Phi(\bar Z, t)^2}_{\text{Field (F)}}-\underbrace{\frac{\lambda}{m}\tilde p \tilde \Phi(\bar Z,t)}_{\text{OF}} - \underbrace{\frac{\lambda}{m} \partial_x\bar\Phi(\bar Z,t)\tilde p\tilde Z}_{\text{OM}}+\underbrace{\frac{\lambda^2}{m}\partial_x\bar \Phi(\bar Z,t)\tilde \Phi(\bar Z,t)\tilde Z-\lambda\dot{\bar q }\partial_x\tilde\Phi(\bar Z,t)\tilde Z}_{\text{MF}}
\label{ham}
\end{align}
We notice that the \textit{mdf} now observes a renormalized oscillation frequency and the scalar field sees a shift coming from the term quadratic in $\Phi$ which is analogous to the diamagnetic term $\sim \frac{e^2}{2mc^2} A^2$ of the minimal coupling Hamiltonian. The bilinear interaction terms represent the coupling between the \textit{idf} and the field (OF), mirror and the \textit{idf} (OM) and mirror and the field (MF) respectively. Physically, the terms that are second order in $\lambda$ arise from the field-field, mirror-mirror and field-mirror couplings mediated via the quantum fluctuations of the \textit{idf}. The terms that are first order in $\lambda$ in the couplings between the {\it idf}-{\it mdf} (OM) and {\it mdf}-field (MF) fluctuations come from the classically driven solutions for the field and the \textit{idf} respectively. Specifically, we note that the MF interaction contains two terms, the first one of which represents the effective mirror-field interaction mediated via the quantum fluctuations of the {\it idf}, while the second one represents that from the classical surface charge currents. Since the conventional approach does not include the fluctuations of this extra quantum degree of freedom, it misses out on the fluctuation-mediated part of the effective mirror-field coupling. As we shall see later, this term becomes dominant in the strong coupling regime. 

We also note that in the absence of a classical drive, the only interaction is between the \textit{idf} and the field (OF)  up to second order. To be able to see an effective mirror-field interaction one needs to include third order terms in the fluctuations as illustrated before.


In the following section we study the above Hamiltonian for the case of a driven single field mode and find the subsequent entanglement dynamics for the mirror CoM and the field, coarse-graining over the internal degrees of freedom.

\section{Mirror-Field Entanglement in the MOF model}
\label{entmf}
Entanglement between a field and a mechanical oscillator has been widely studied in cavity optomechanical setups in several contexts \cite{Asp13}--\cite{YC}, with the essential mirror-field coupling mechanism being the radiation pressure wherein the field exerts a force on the mirror center of mass by means of photon-momentum transfer and observes a phase shift proportional to the mirror displacement in turn. We now look at the entanglement generation from a microscopic perspective as described by the MOF model, considering only a single mode of the scalar field in our model as in the usual cavity optomechanical setups to deduce some key physical features of the mirror-field entanglement that arise from the inclusion of the {\it idf}.


We first simplify the Hamiltonian \eqref{ham} for the case of a single field mode that is being externally driven to look at the dynamics of the coupled MOF system and then coarse-grain the \textit{idf} to find the sought after mirror-field entanglement. 
Consider the scalar field in a region of length L (assuming L approaches infinity), the field fluctuations can then be written as the sum of all discrete modes of the cavity of length L as $\tilde \Phi\bkt{x,t} = \sum_{n} \sqrt{\frac{\hbar}{2\omega_n\epsilon_0L}}\bkt{\tilde a_n e^{ik_nx}+\tilde a^\dagger_n e^{-ik_nx}}$, with $\tilde a^\dagger_n$ and $\tilde a_n$ representing the creation and annihilation operators for the n$^\text{th}$ field mode. We pick a single field mode at frequency $\omega$ interacting with the point mirror at the origin $\bkt{\bar Z =0}$ assuming that the center of mass motion is in the sub-wavelength regime as before. 
\begin{align}
\tilde{\Phi}_\omega\bkt{x,t} = \sqrt{\frac{\hbar}{2\omega\epsilon_0L}}\bkt{\tilde a_\omega e^{ikx}+\tilde a^\dagger_\omega e^{-ikx}}
\end{align}

The above expression represents the fluctuations of the free field without any imposed boundary conditions unlike the standard treatment where the quantum fluctuations follow the mode functions of the classical field (see \cite{Kempf} for example).  In the steady state, the strength of the field fluctuations would be determined by the boundary conditions as they emerge from the \textit{idf}-field interaction self-consistently.

For a single field mode, we rewrite the free Hamiltonian part in \eqref{ham} as 
\begin{align}
\tilde{H}_{free}\equiv&\underbrace{\frac{\tilde P^2}{2M} + \frac{1}{2} M\mho'^2\tilde Z^2}_{\tilde H_M}+\underbrace{\hbar \Omega\bkt{\tilde b^\dagger\tilde b+\frac{1}{2}}}_{\tilde H_O}+\underbrace{ \hbar\bkt{ \omega+\frac{\lambda^2}{2m\omega\epsilon_0L}}\bkt{\tilde a_\omega^\dagger \tilde a_\omega +\frac{1}{2}}+\frac{\lambda^2}{4m\omega\epsilon_0L}\bkt{\bkt{\tilde a_\omega}^2+\bkt{\tilde {a}_\omega^\dagger}^2 }}_{\tilde H_F}
\label{hf}
\end{align}
where we have redefined the dynamical variables associated with the \textit{idf} in terms of the creation annihilation operators $\{\tilde b ^\dagger, \tilde b\}$ as $\tilde q =\sqrt{\frac{\hbar}{2m\Omega}}\bkt{\tilde b +\tilde b^\dagger }$ and $\tilde p =-i \sqrt{\frac{\hbar m \Omega}{2}} \bkt{\tilde b -\tilde b^\dagger}$. The renormalized mechanical frequency is defined as $\mho'^2\equiv\mho^2+\frac{\lambda^2}{mM}\bkt{\partial_x\bar\Phi(\bar Z,t)}^2$. As mentioned in the previous section, the correction term $\bkt{\frac{\lambda^2}{mM}\bkt{\partial_x\bar\Phi(\bar Z,t)}^2}$ contains two contributions - a time dependent part oscillating at a frequency $\sim 2\omega$ and a time-independent part. In the rotating wave approximation (RWA) the time dependent term can be neglected. However, if the field mode was resonant with the \textit{mdf}, one would see parametric amplification of the mirror center of mass motion \cite{Nori}. For the free field part we notice that the interaction leads to an energy correction $\omega\rightarrow\omega+\lambda^2/(2m\omega\epsilon_0L)$ that is second order in $\lambda$, this corresponds to the shift coming from the diamagnetic contribution $\bkt{\sim \frac{e^2}{2mc^2}A^2}$ for the EM case as indicated in the previous section. This also leads to the fast oscillating terms for the free field ($\sim2\omega$), which correspond to the photon-pair production and annihilation as in the case of dynamical Casimir effect \cite{Nori, Moore, FD, Wilson}.
Moving to the interaction picture with respect to $\tilde H_0 = \tilde H_O +\tilde H_F$ to eliminate the fast dynamics of the system and invoking RWA, we write the interaction Hamiltonian in a simplified form as
\begin{align}
\tilde H_{int} \equiv \hbar\bkt{\alpha_{OF}\boldsymbol{b}^\dagger\boldsymbol{a}e^{-i\Delta t}+\alpha_{OF}^\ast\boldsymbol{b}\boldsymbol{a}^\dagger e^{i\Delta t}}+\hbar\bkt{\alpha_{OM}\boldsymbol{b} e^{i\Delta t}+\alpha_{OM}^\ast\boldsymbol{b}^\dagger e^{-i\Delta t}}\bkt{\tilde c +\tilde c^\dagger} +\hbar\bkt{\alpha_{MF}\boldsymbol{a}+\alpha_{MF}^\ast\boldsymbol{a}^\dagger}\bkt{\tilde c +\tilde c^\dagger}
\label{Hint}
\end{align}
Here we have defined the operators in the interaction picture as $\{\boldsymbol a,\boldsymbol a^\dagger \}\equiv \{\tilde a_\omega e^{i\omega t},\tilde a^\dagger_\omega e^{-i\omega t}\}$ and $\{\boldsymbol b,\boldsymbol b^\dagger \}\equiv \{\tilde be^{i\Omega t},\tilde b^\dagger e^{-i\Omega t}\}$ and the detuning $\Delta \equiv \omega-\Omega$ represents the detuning between the field and the \textit{idf}. The operators $\{\tilde c, \tilde c^\dagger\}$ correspond to the creation and annihilation operators for the phononic excitations of the \textit{mdf}, with $\tilde {Z} = \sqrt{\frac{\hbar}{2 M\mho'}}\bkt{\tilde c +\tilde c^\dagger}\equiv \sqrt{\frac{\hbar}{ M\mho'}}\boldsymbol{Z}$ and $\tilde {P} = -i\sqrt{\frac{\hbar M\mho'}{2}}\bkt{\tilde c -\tilde c^\dagger}\equiv \sqrt{\hbar M\mho'}\boldsymbol{P}$. The operators $\boldsymbol Z$ and $\boldsymbol P$ are the dimensionless position and momentum fluctuations for the mirror center of mass. In moving to the interaction picture we have ignored the second order correction terms $\bkt{\sim\lambda^2/m}$ in the free field Hamiltonian $\tilde H_F$.

The coefficients $\alpha_{ij}$s represent the effective bilinear coupling strengths between the single excitations of the three subsystems with
\begin{align}
\label{alphaof}
\alpha_{OF} &\equiv - \frac{i\lambda}{2}\sqrt{\frac{\Omega}{m\omega\epsilon_0 L}}\\
\label{alphaom}
\alpha_{OM} &\equiv  \frac{\Omega \Phi_0\lambda}{2c}\sqrt{\frac{ \Omega}{mM\mho'}}\\
\label{alphamf}
\alpha_{MF}&\equiv \frac{\Omega \Phi_0}{2c} \sqrt{\frac{1}{M\mho'\omega\epsilon_0L}}\bkt{-\frac{i\lambda^2}{m}+2\epsilon_0c\omega R^\ast(\omega)}  = \frac{\Omega A_0}{L} \sqrt{\frac{\hbar}{2M\mho'}}\bkt{-\frac{i\lambda^2}{2mc\omega\epsilon_0}+ R^\ast(\omega)}
\end{align}
where we have defined the dimensionless field amplitude $A_0 \equiv \Phi_0 /\sqrt{\frac{\hbar}{2\omega\epsilon_0 L}}$. It can then be seen from \eqref{alphamf} that for a perfectly reflecting mirror $\bkt{R^\ast(\omega)\rightarrow-1}$ the second term in the effective coupling strength $\alpha_{MF}$ between the {\it mdf} and the field is the same as what one finds from standard boundary condition approach (see Appendix \ref{compareBC}).

\begin{figure}[ht]
\subfloat[]{\includegraphics[width=3.5 in]{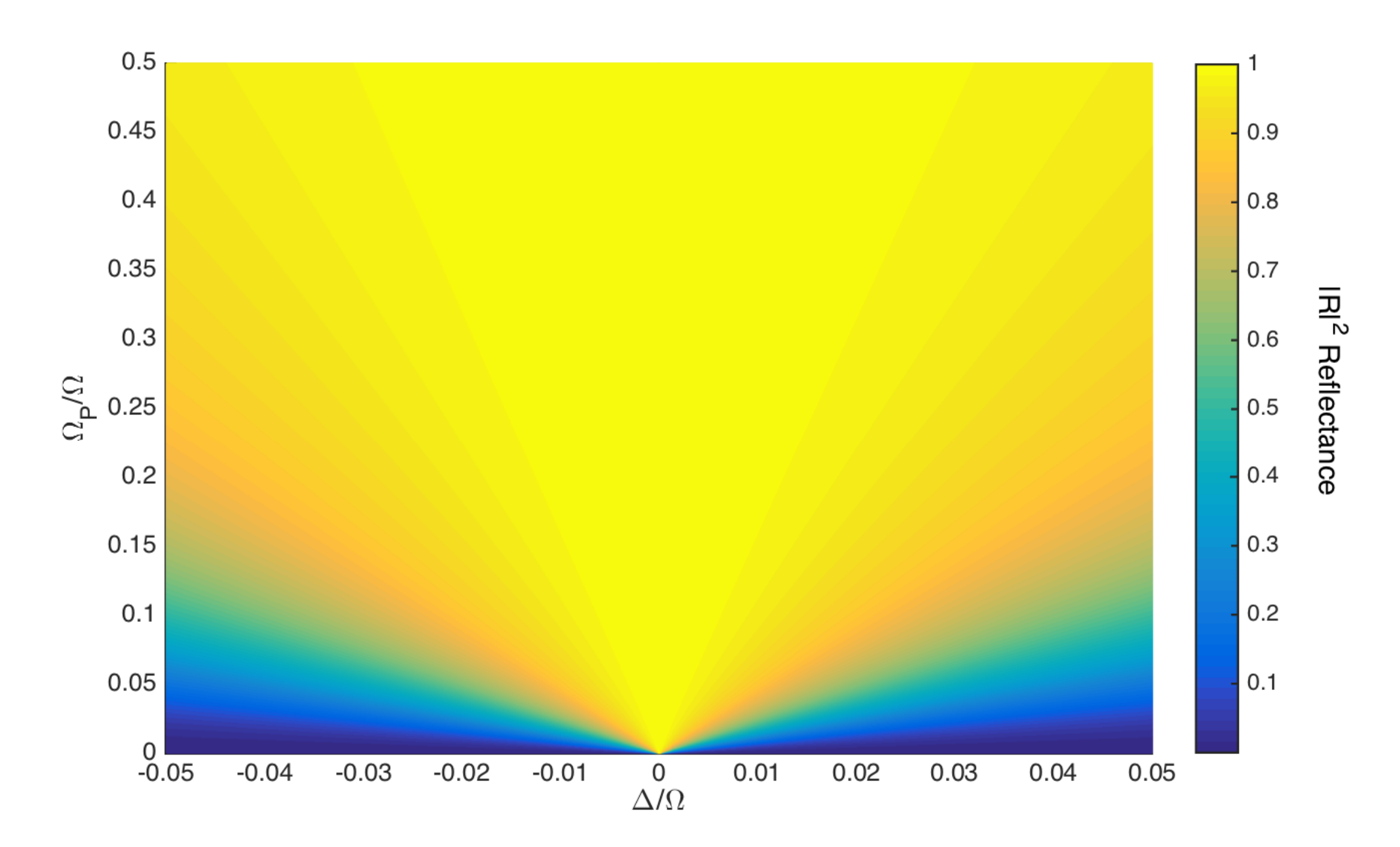}}
\subfloat[]{\includegraphics[width=3.5 in]{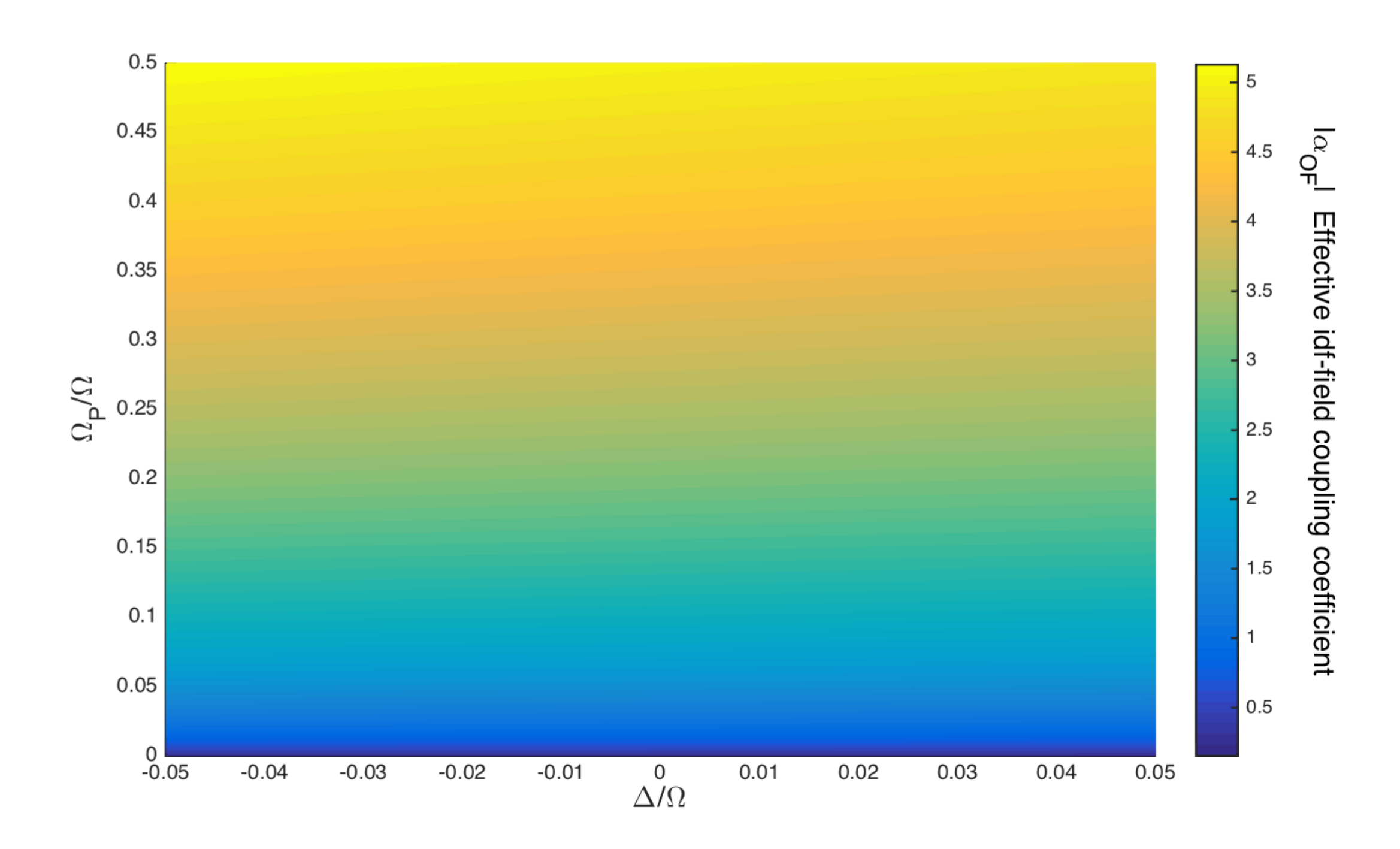}}\\
\subfloat[]{\includegraphics[width=3.5 in]{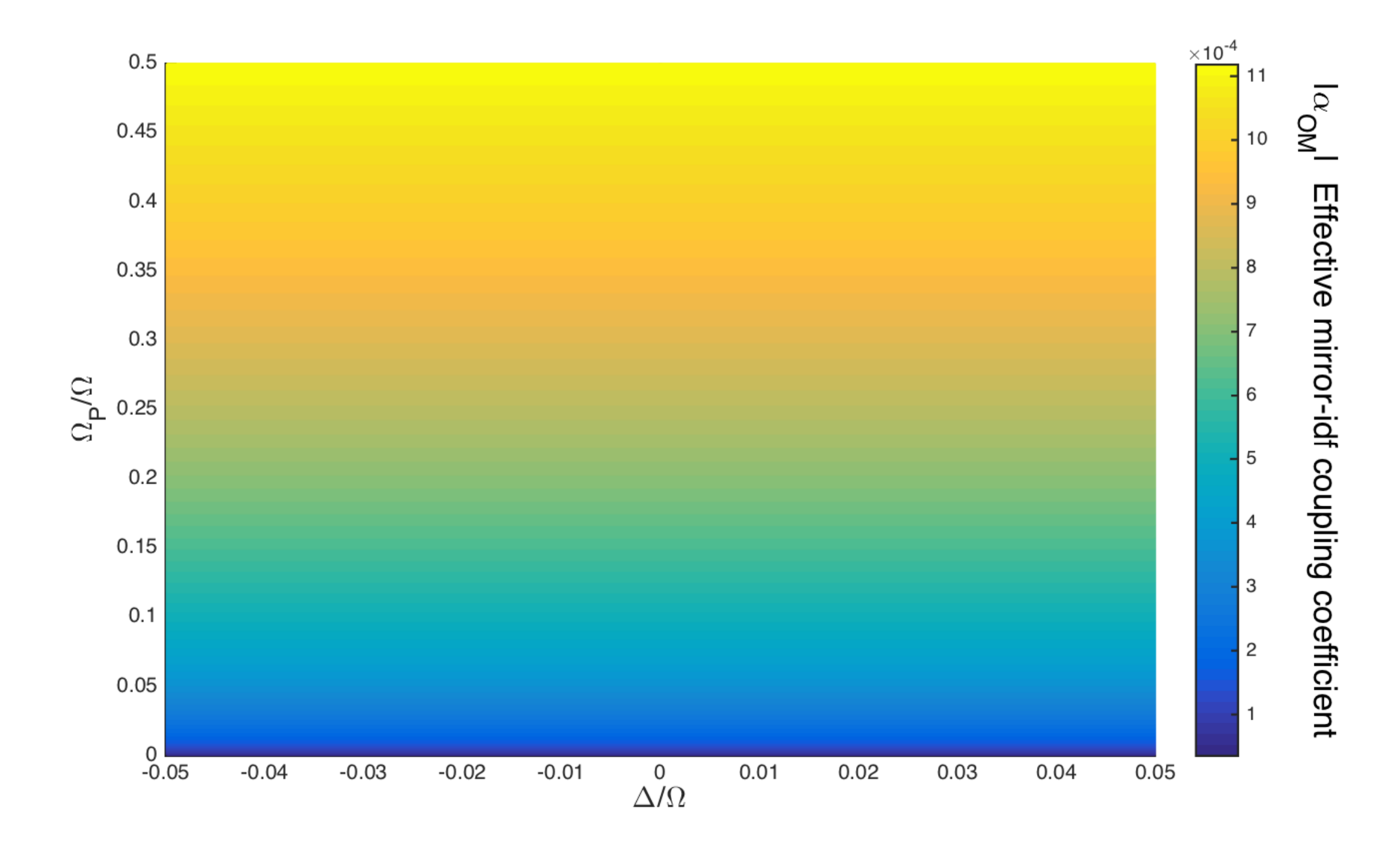}}
\subfloat[]{\includegraphics[width=3.5 in]{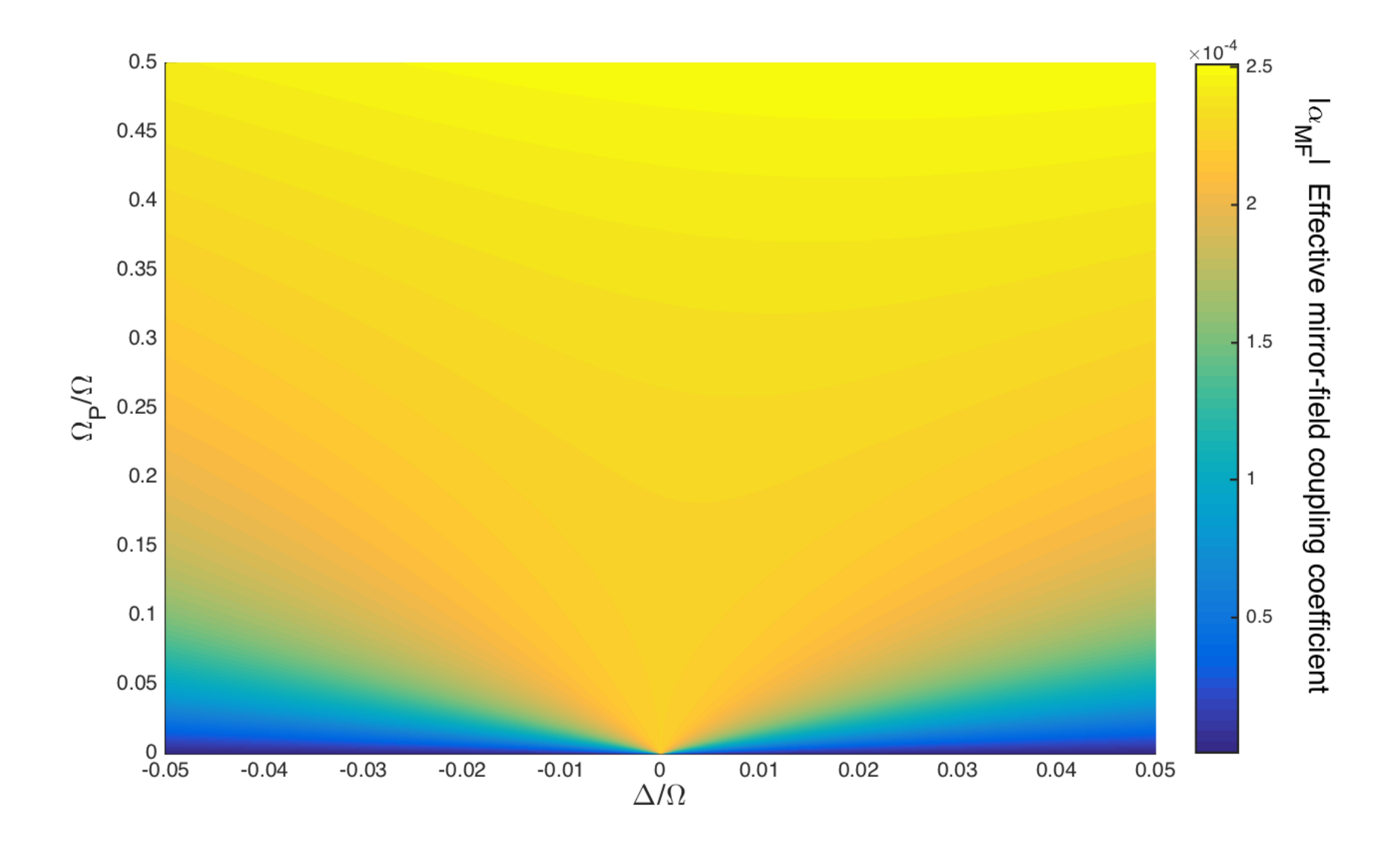}}\\
\caption{(a) Reflectance as a function of the dimensionless parameter $\Omega_P/\Omega$ (ratio of the plasma frequency to the {\it idf}'s natural frequency) and the {\it idf}-field detuning $\Delta/\Omega$. It can be seen that for weaker coupling corresponding to $\Omega_P/\Omega\ll1$, the reflection spectrum has a sharper resonance. The effective bilinear coupling strengths for both (b)\textit{idf}-field $\bkt{\alpha_{OF}}$ and (c)\textit{idf}-\textit{mdf} $\bkt{\alpha_{OM}}$ increase with increasing plasma frequency as $\sim\sqrt{\Omega_P}$.  (d)The effective \textit{mdf}-field coupling coefficient $\bkt{\alpha_{MF}}$ in the weak coupling limit is largely determined by the reflection coefficient, while for strong coupling the fluctuation mediated part becomes relevant, as can be seen from \eqref{alphamf}.}
\label{coupling}
\end{figure}

We note that all the effective coupling strengths contain the \textit{idf} mass and charge parameters in the combination $\sim\lambda^2/m$ which corresponds to the plasma frequency $\Omega_P (\equiv\lambda^2/(2m\epsilon_0 c ))$, meaning that one can deduce all the effective single excitation couplings ($\alpha_{ij}$s) from the two parameters that also completely characterize the reflection spectrum, $\Omega$ and $\Omega_P$ as defined in section \ref{OP}. Thus given the reflection spectrum of a mirror, one can find the parameters $\Omega_P$ and $\Omega$, knowing which one arrives at the various effective coupling strengths. Fig.\ref{coupling} shows the dependence of the reflection coefficient and these effective couplings  on the dimensionless plasma frequency $\bkt{\Omega_P/\Omega}$ and detuning $\bkt{\Delta/\Omega}$.

It can also be observed from \eqref{alphaof}--\eqref{alphamf} that the coupling strengths increase as the original \textit{idf}-field coupling $\lambda$ increases and decrease as the \textit{idf} mass $m$ increases, meaning that a "lighter" \textit{idf} leads to stronger effective coupling strengths. Also, a heavier mirror CoM couples more weakly to the \textit{idf} and the field. The effective \textit{idf}-field coupling $\alpha_{OF}$ is independent of the driving field amplitude $\Phi_0$ as expected, since as one sets the drive amplitude to zero it can be seen that there is no mirror-field and \textit{idf}-mirror interaction in second order except the {\it idf}-field coupling from the direct interaction. To be able to see any coupling and hence entanglement between the mirror and the field in that case one needs to include the higher order terms as was discussed before.

We also note here that in the weak coupling regime where $\Omega_P\ll1$, the mirror reflectivity and the effective mirror-field coupling strength $\alpha_{MF}$ as a function of the  \textit{idf}-field detuning peaks sharply at resonance $(\Delta=0)$  as seen from Fig.\ref{coupling}(a) and Fig.\ref{coupling} (d). The field amplitude and detuning with respect to the \textit{idf} change the coupling strengths appreciably. While in the standard treatment of mirror-field interactions via boundary conditions it is common to study the effect of the field intensity on the mirror-field coupling, we highlight that including the presence of \textit{idf} lets us see the effect of the field-\textit{idf} detuning on the mirror-field interaction, allowing us to probe the effective coupling strength as a function of the reflection properties of the mirror. 

As noted before, the two terms in the effective mirror-field coupling  $\alpha_{MF}$ denote the interaction mediated via the quantum fluctuations of the $\idf$ and its classical amplitude respectively. The strong-coupling limit, where one would expect to see non-Markovian dynamics is also where the contribution from the {\it idf} fluctuations becomes substantial, as seen from the first term in \eqref{alphamf}.

Now we use the interaction Hamiltonian \eqref{Hint} to write the equations of motion in terms of the coupling constants $\alpha_{ij}$s as
\begin{align}
\label{qeom1}
\der{\boldsymbol{Z}}{t} =& \mho'\boldsymbol{P}\\
\label{qeom2}
\der{\boldsymbol{P}}{t} =& -\mho'\boldsymbol{Z}-2\bkt{\re\alpha_{OM}\mathbf{q}-\im\alpha_{OM}\boldsymbol p} -2\bkt{\re\alpha_{MF}\mathbf{\Phi}-\im\alpha_{MF}\boldsymbol{\Pi}}-\gamma\boldsymbol P +\tilde \xi\\
\label{qeom3}
\der{\boldsymbol{q}}{t} =& \Delta\boldsymbol{p}-\vert\alpha_{OF}\vert\boldsymbol{\Phi}-2\im\alpha_{OM}\boldsymbol Z\\
\label{qeom4}
\der{\boldsymbol{p}}{t} =& -\Delta\boldsymbol{q}-\vert\alpha_{OF}\vert \boldsymbol{\Pi}-2\re\alpha_{OM}\boldsymbol{Z}\\
\label{qeom5}
\der{\boldsymbol{\Phi}}{t} =& \vert\alpha_{OF}\vert\boldsymbol{q}-2\im\alpha_{MF}\boldsymbol{Z}\\
\label{qeom6}
\der{\boldsymbol{\Pi}}{t} =& \vert\alpha_{OF}\vert\boldsymbol{p}-2\re\alpha_{MF}\boldsymbol{Z}
\end{align}
wherein we have redefined the slow moving dimensionless \textit{idf} and the field quadratures as $\boldsymbol{q}\equiv\frac{\boldsymbol{b}e^{i\Delta t}+\boldsymbol{b}^\dagger e^{-i\Delta t}}{\sqrt 2}$, $\bold{p}\equiv-i\frac{\boldsymbol{b}e^{i\Delta t}-\boldsymbol{b}^\dagger e^{-i\Delta t}}{\sqrt{2}}$, $\bold{\Phi}\equiv \frac{\boldsymbol{a}+\boldsymbol{a}^\dagger}{\sqrt{2}}$ and $\bold{\Pi}\equiv -i\frac{\boldsymbol{a}-\boldsymbol{a}^\dagger}{\sqrt 2}$. Also, to account for the fluctuation-dissipation mechanism for the mirror center of mass resulting from its coupling to the thermal bath, we have introduced the mechanical damping $\gamma$ and noise $\tilde \xi$ for the mirror. In the high temperature limit, the correlation function of the noise is given as $\avg{\tilde \xi\bkt{t}\tilde \xi\bkt{t'}} = 2M\gamma k_B T \delta(t-t')$, with T as the temperature of the thermal bath.

\begin{figure}[ht]
\includegraphics[width = 4 in]{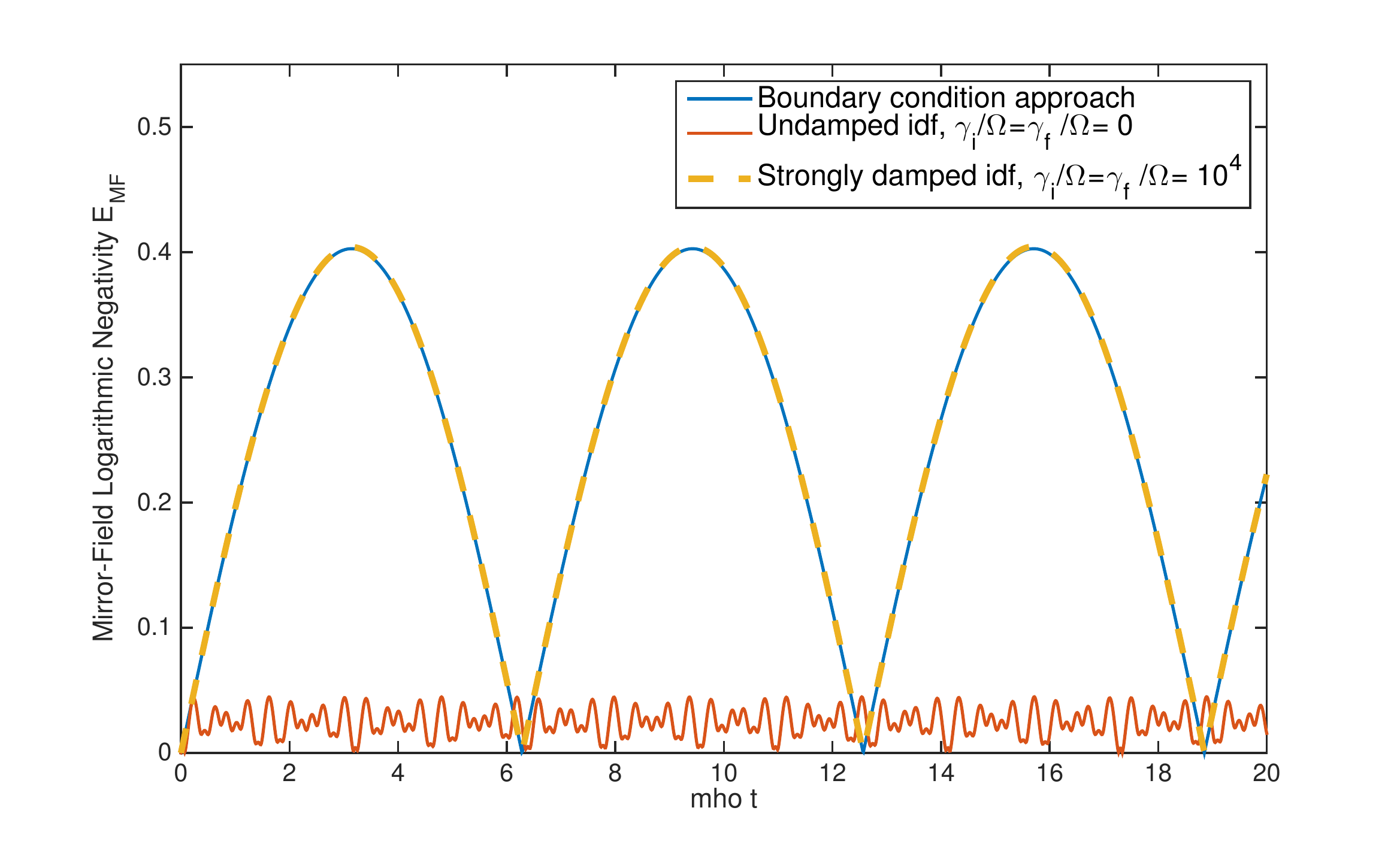}
\caption{Evolution of mirror-field entanglement as measured by the logarithmic negativity $E_{MF}$ (see Appendix \ref{LN} for definition) as obtained from the boundary condition approach (Appendix \ref{compareBC}) and the coupled MOF dynamics. We find that for an isolated {\it idf} the time scale for entanglement is largely determined by the effective {\it idf}-field coupling ($\alpha_{OF}$). The two approaches concur in the weak coupling limit for a strongly damped {\it idf}. The parameters values, in units where c=1, $\hbar=1$, used here are $m = 0.001$, $\Omega=100$, $M = 10$, $\mho =0.1$, $\Omega_P = 5$, $A_0 = 10^{-4}$ and $T = 1000$. The effective {\it idf}-field coupling strength $\vert\alpha_{OF}\vert/\mho \approx 16$.}
\label{comp}
\end{figure}

At this point one can make a crucial observation that if we assume weak-coupling such that $\Omega_p\ll\Omega$, the fluctuation mediated part of the effective mirror-field coupling is negligible or $\alpha_{MF}\approx \frac{\Omega A_0}{L} \sqrt{\frac{\hbar}{2M\mho'}} R^\ast(\omega)$. Additionally, if we consider the field detuning to be small enough such that $\Delta\ll\Omega_p$, the reflection coefficient $R^\ast(\omega)\approx1$ leading to a nearly perfectly reflecting mirror. In this parameter regime, it can be seen that the mirror-field dynamics from \eqref{qeom1}-\eqref{qeom6} found from the MOF model reduces to that from the conventional boundary condition approach (see Appendix \ref{compareBC} for further details), provided that the {\it idf} contribution is negligibly small. To ensure which we consider that the {\it idf} is coupled to the continuum of field modes with a coupling of the form $\dot{q}\Phi_i$, where $\Phi_i$ represents the i$^{th}$ field mode, leading to a damping coefficient $\gamma_f$. Also, to mimic the scattering of surface charges by lattice ions of the mirror, we introduce a dissipative bath of internal degrees of freedom such that each bath oscillator is coupled to the {\it idf} with a coupling of the form $q \cdot q_i$, where $q_i$ represents the position variable for the i$^{th}$ bath oscillator, giving an effective damping coefficient of $\gamma_i$ for the {\it idf}. Using separation of time scales, we find the steady state {\it idf} amplitudes as $\boldsymbol{q}_{st} = -\frac{\gamma_i \hat C_1 +\Delta \hat C_2}{\Delta^2+\gamma_i\gamma_f}$ and $\boldsymbol{p}_{st} = \frac{\gamma_f \hat C_2 -\Delta \hat C_1}{\Delta^2+\gamma_i\gamma_f}$, where the operators $\hat C_i$s stand for $\hat C_1 \equiv |\alpha_{OF}| \boldsymbol{\Phi} +2 \mathrm{Im} \alpha_{OM} \boldsymbol {Z}+\hat \xi_f$ and
$\hat C_2 \equiv |\alpha_{OF}| \boldsymbol{\Pi} +2 \mathrm{Re} \alpha_{OM} \boldsymbol {Z}+\hat \xi_i$. Now for the case of near perfect reflection since the detuning $\Delta$ is small, for the steady state amplitudes to vanish, we must have $\gamma_{i,f}\gg\Delta$. It can then be seen that the dynamics obtained from the two approaches agree perfectly with each other as shown in Fig. \ref{comp}.
%

Now going back to the case of an isolated {\it idf}, we can identify the radiation pressure force from \eqref{qeom2} as 
\begin{align}
\tilde{F}_{rad}\equiv-2\bkt{\re\alpha_{OM}\bold{q}-\im\alpha_{OM}\bold p}-2\bkt{\re\alpha_{MF}\bold{\Phi}-\im\alpha_{MF}\bold{\Pi}}
\label{rp3}
\end{align}
We can see that the linearized radiation pressure force depends on both the fluctuations of the \textit{idf} and the field variables and is generally dependent on the \textit{idf} parameters. Hence as long as the {\it idf} fluctuations are non-vanishing, the radiation pressure shot noise is determined by the shot noise of both the field and the \textit{idf}, meaning that in order to go below the standard quantum limit for the radiation pressure force one needs to take into consideration the squeezing of the \textit{idf} quadratures in addition to those of the field \cite{YCRev, Purdy13}. 

From \eqref{qeom1}--\eqref{qeom6} we write the solutions to the equations of motion for the {\it idf} and the field variables as
\begin{align}
\bold{q}(t)& = \bold{q}_h(t)+\int_0^t\dd t'G_o(t-t')\bkt{-\Delta\vert\alpha_{OF}\vert\bold{\Pi}(t')-\Delta\alpha_{OM}\boldsymbol{Z}(t')+2\vert\alpha_{OF}\vert\im\alpha_{MF}\boldsymbol{Z}(t')}\\
\bold{p}(t)& = \bold{p}_h(t)+\int_0^t\dd t'G_o(t-t')\bkt{\Delta\vert\alpha_{OF}\vert\bold{\Phi}(t')+2\vert\alpha_{OF}\vert\re\alpha_{MF}\bold{Z}(t')-\alpha_{OM}\bold{P}(t')}\\
\bold{\Phi}(t)& = \bold{\Phi}_h(t)+\int_0^t\dd t'G_f(t-t')\bkt{\Delta\vert\alpha_{OF}\vert\bold{p}(t')-2\im\alpha_{MF}\bold{P}(t')}\\
\bold{\Pi}(t) &= \bold{\Pi}_h(t)+\int_0^t\dd t' G_f(t-t')\bkt{-\Delta\vert\alpha_{OF}\vert\bold{q}(t')-\vert\alpha_{OF}\vert\alpha_{OM}\bold{Z}(t')-2\re\alpha_{MF}\bold{P}(t')}
\label{sol}
\end{align}
Here we have defined the \textit{idf} and the field Green's functions as $G_O(t)\equiv\frac{\sin(\sqrt{\vert\alpha_{OF}\vert^2+\Delta^2}t)}{\sqrt{\vert\alpha_{OF}\vert^2+\Delta^2}}$ and $G_f(t)\equiv\frac{\sin(\vert\alpha_{OF}\vert t)}{\vert\alpha_{OF}\vert}$ and the homogeneous solutions as $\{\bold{q}_h,\bold{p}_h,\bold{\Phi}_h,\bold{\Pi}_h\}$. It can be seen that the frequency of oscillations for the slow moving \textit{idf} variables is $\Omega_{idf}\equiv\sqrt{\abs{\alpha_{OF}^2}+\Delta^2}$ and that for the slow moving field variables is $\Omega_{f}\equiv \abs{\alpha_{OF}}$.
In the steady state limit, we can use these solutions to rewrite the equation of motion for the late time mirror CoM dynamics as
\begin{align}
&M\bkt{\partial_t^2+\mho'^2}\bold{Z}(t) +\gamma\der{\bold{Z(t)}}{t} +\int_0^t\dd t'G_o(t-t')\bkt{-\Delta\alpha_{OM}^2+2\alpha_{OF}\im\alpha_{MF}\alpha_{OM}}\bold{Z}(t')=\nonumber\\
& \int_0^t\dd t'G_o(t-t')\bkt{2\alpha_{OM}\Delta\vert\alpha_{OF}\vert\bold{\Pi}(t')} +\int_0^t\dd t' G_f(t-t')\bkt{2\re\alpha_{MF}\Delta\vert\alpha_{OF}\vert\bold{p}(t')+2\im\alpha_{MF}\Delta\vert\alpha_{OF}\vert\bold{q}(t')}+\tilde \xi&
 \label{solZ}
\end{align}
On the left side one can identify the two terms in the integral as the retarded influence of mirror-idf-idf-mirror interaction and the mirror-idf-field-mirror interaction respectively. The first term on the right side denotes the mirror being driven by the {\it idf}-influenced field and the second term stands for the mirror being driven by the field-influenced {\it idf}. We can see that in the absence of any detuning the CoM motion is only driven by the thermal noise term.

One can find the normal modes of the system from \eqref{qeom1}--\eqref{qeom6} and their time evolution to obtain the 6x6 dimensional covariance matrix of the coupled MOF system numerically. We define the MOF covariance matrix $\mathcal V_{MOF}$ as
\begin{align}
\mathcal V_{MOF} = \left(\begin{array}{ccc}
\bf {V}_{MM} &\bf V_{MF} &\bf V_{OM}\\
\bf {V}^T_{MF} & \bf V_{FF} &\bf V_{OF}\\
\bf V^T_{OM} & \bf V^T_{OF} &\bf V_{OO}
\end{array}\right)
\end{align}
where the on-diagonal sub-matrix $\bf V_{kk}$ stands for the covariance matrix of the k$^{th}$ reduced subsystem, defined as $\bkt{\bf V_{kk}}_{ij} \equiv\frac{1}{2}\avg{\{X^{(k)}_i,X^{(k)}_j\}}$, with ${X}_i^{(k)}$ and ${X}_j^{(k)}$ representing the i and j quadratures corresponding to the position and momentum variables of the k$^{th}$ reduced subsystem, more explicitly  $\vect X^{(k)}\equiv\{\tilde x^{(k)},\tilde p^{(k)}\} $. Here, $\{ \mathcal O_1, \mathcal O_2\}$ denotes the anti-commutator between the operators $\mathcal O_1$ and $ \mathcal O_2$. The off-diagonal sub-matrix $\bf V_{kl}$ consists of the correlations between the k$^{th}$ and the l\th subsystems, such that $\bkt{\bf V_{kl}}_{ij} \equiv\frac{1}{2}\avg{\{X^{(k)}_i,X^{(l)}_j\}}$, where the i and j quadrature components belong to different subsystems.

We then choose to look at the part of the covariance matrix that represents the mirror and field reduced covariance matrices and correlations, that is,
\begin{align}
\mathcal V_{MF}= \left(\begin{array}{cc}
\bf {V}_M &\bf V_{MF} \\
\bf {V}^T_{MF} & \bf V_{F}
\end{array}\right)
\label{cmf}
\end{align}
and find the logarithmic negativity $E_{N}^{MF}$ as obtained based on the positive partial transpose (PPT) criteria for determining separability (see Appendix \ref{LN} for details). It can be shown that calculating the MF entanglement from the sub covariance matrix $\mathcal V_{MF}$ is equivalent to coarse-graining over the internal degree of freedom and then finding the MF entanglement.
\begin{figure}[ht]
\includegraphics[width=4 in]{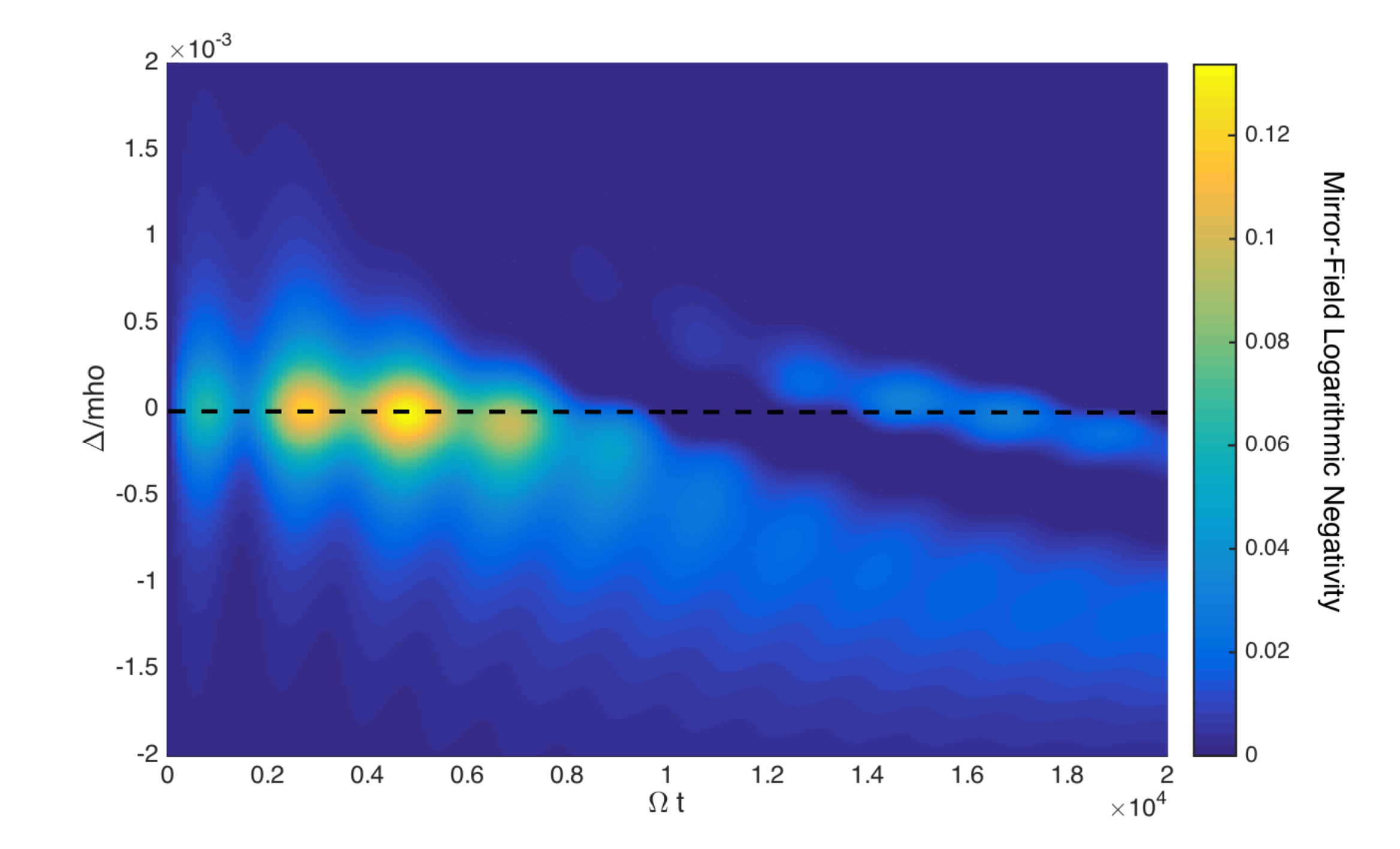}
\caption{Mirror-Field entanglement given by the logarithmic negativity for an undamped {\it idf} as a function of the dimensionless \textit{idf}-field detuning ($\Delta/\mho$) and dimensionless time ($\Omega t$). We observe that the entanglement peaks for a resonant {\it idf}-field interaction at ($\Delta/\mho =0$) and the oscillation time scales are determined by the effective {\it idf}-field coupling ($|\alpha_{OF}|$). The parameters values, in units where c=1, $\hbar=1$, used here are $m = 0.001$, $\Omega=100$, $M = 10$, $\mho =0.1$, $\Omega_P = 0.05$, $A_0 = 10^{-4}$ and $T = 1000$.}
\label{MFent}
\end{figure}


As was discussed before, at the \textit{idf}-field resonance ($\Delta \rightarrow0$) the reflection coefficient and hence the effective mirror-field coupling strength goes to its maximum value (See Fig.\ref{coupling}(a) and Fig. \ref{coupling}(d)). As a result, we observe in Fig.\ref{MFent}  that there is a peak in the mirror-field entanglement near \textit{idf}-field resonance. As was emphasized before, this effect is not considered in the standard treatment of optomechanical interactions since the internal degree is coarse-grained over a priori to arrive at the boundary conditions for the field. This effect is more pronounced in the weak-coupling regime where the reflection coefficient has a sharper peak at resonance as seen from Fig.\ref{coupling}(a) and Fig.\ref{coupling}(d). It can also be observed that for $\Delta/\mho = -1$ or equivalently $\Omega = \omega+\mho$, the entanglement is sustained for longer times. Physically, this pertains to the process wherein a field photon and a mirror phonon combine to give a single {\it idf} excitation (or vice versa), corresponding to the two-mode squeezing Hamiltonian which then entangles the field and the mirror modes as a result of the interaction. Such an observation had also been made in \cite{Vitali07} for the case of a cavity driven with a red detuned drive in the sideband resolved regime where it was shown that the steady state entanglement goes to a maximum when the cavity-drive detuning was equal to the mechanical oscillation frequency. Drawing an analogy between the two cases, we find that cavity resonance for the usual cavity optomechanical setups is similar to the {\it idf} in the MOF model in that they both mediate the interaction between the mechanical motion of the mirror and the external field. Also, comparing with other existing results on mirror-field entanglement such as \cite{YC} we note that in the present setup in the absence of a continuum of field modes our results do not indicate the existence of a steady-state entanglement at high temperatures.

While the parameter values chosen here may pertain to a narrow parameter regime corresponding to weak coupling and isolation from environment, we have illustrated that there is a significant effect of the \textit{idf} parameters on the mirror-field entanglement. We discuss our results further and conclude in the following section.

\section {Discussion}

The foremost theme in our analysis is to highlight the significance of the internal degrees of freedom of a mirror that play the role of the essential intermediary when it comes to studying the interaction between a quantum field and the mirror's mechanical motion. We illustrate how a microscopic model of quantum optomechanics, such as the MOF model proposed by Galley, Behunin and Hu \cite{CR} is a physically more complete and intuitive description for optomechanical interactions, in that not only can it agreeably reproduce the known optomechanical properties both in the classical and quantum regimes, it also elucidates new physical aspects which are not accounted for in the general description of optomechanical interactions via radiation pressure coupling. Specifically looking at the quantum entanglement between the mirror's mechanical motion and the field, we find that there is a significant and even a critical role played by the internal degree of freedom in certain parameter regimes as it can act as a means to coherently transfer correlations between the field and the mechanical degree of freedom.

The MOF model allows us to go beyond the usual disjoint treatment of mirror-field interactions wherein one imposes boundary conditions on the field and treats the mechanical effects of the field arising from the radiation pressure force separately to attain a self-consistent depiction where we see both the radiation pressure (section \ref{RP}) and the boundary conditions (section \ref{OP}) emerge from a physically motivated microscopic interaction. The new key aspects that arise from this self-consistent treatment of the mirror-field interaction can be summarized as follows.

\begin{itemize}

\item \textit{Coherent transfer of excitations} -- We show that the conventional boundary condition approach arises as the limiting case of the MOF model where the quantum fluctuations of the {\it idf} are lost to the bath. As illustrated in Fig.\ref{comp}, isolating the {\it idf} from the environment can provide an additional channel for coherent transfer of mirror-field correlations. Since the {\it idf}-field dynamics is at a much faster time scale as compared to the center of mass motion, for an undamped {\it idf} we observe much faster time scales for the entanglement dynamics determined by the effective {\it idf}-field coupling $\bkt{\alpha_{OF}}$ rather than those from the conventional radiation pressure coupling $\bkt{\sim\mho}$.

\item \textit{Fully dynamical description} -- For relativistically moving mirrors, as in the case of dynamical Casimir effect \cite{Wilson}, applying boundary conditions is an inadequate description of the dynamics since in the time scales over which internal degrees and the field reach a steady state thereby leading to an effective boundary condition, the mirror center of mass moves appreciably enough to affect their interaction. In cases where the timescales of the mechanical motion and the field-internal degree of freedom interaction are close to each other, including the internal degree of freedom becomes relevant as the only means to capture the coupled dynamical interplay of the three subsystems.

\item \textit{Field frequency shift} -- In the MOF description, we observe an additional shift to the field frequency from its second order interaction with the \textit{idf} as seen from \eqref{ham} and \eqref{hf}, a feature that is not accounted for in the boundary condition treatment. Such a diamagnetic term contribution can be significant in the strong coupling regimes, even leading to change in the radiation pressure force from attractive to repulsive as has been studied in \cite{Kempf}.

\item \textit{Radiation pressure shot noise} -- We also observe that for an undamped {\it idf} the radiation pressure force is determined not only by the quantum fluctuations of the field but also those of the {\it idf} as suggested by \eqref{rp3}. While in the steady state limit the strength of these fluctuations is largely determined by the boundary conditions, in the early time limit the \textit{idf} being an independent quantum degree of freedom its quantum fluctuations would influence the radiation pressure shot noise as well. 

\end{itemize}

%

As a result of including an extra quantum degree of freedom one would naturally expect there to be a difference in the quantum correlations of the field and the mirror's mechanical motion. We show that in the parameter regimes where the {\it idf} is isolated from the environment and for strong coupling, the role of the {\it idf} is more pronounced. This can be seen from the effective mirror-field interaction strength that is determined strongly by the \textit{idf}-field resonance condition, with the effective bilinear CoM-field coupling going to a maximum at resonance as in \eqref{alphamf}. The dependence of the optomechanical interaction on the \textit{idf}-field resonance leading to an enhanced mirror-field entanglement is something that can not be captured in the boundary condition treatment of optomechanical interactions. This is seen in the plot for MF entanglement as a function of detuning $\Delta$ in Fig.\ref{MFent}. 

Also, the time scales for all dynamics, including that of the mirror-field entanglement is largely determined by the effective \textit{idf}-field coupling $\alpha_{OF}$ and the \textit{idf}-field detuning $\Delta$ as seen from \eqref{sol}. One can make the same observation from Fig.\ref{comp} that for an {\it idf }isolated from the environment, the entanglement dynamics are at a much faster timescale $(\alpha_{OF})$ as compared to the boundary condition approach $(\mho)$.

Analogous to the case of optomechanical entanglement in a cavity setup, we find that if the field is red-detuned with respect to the cavity frequency such that one facilitates the two-mode squeezing interaction between the field and the mechanical mode, the entanglement is sustained for longer times as can be seen from Fig. \ref{MFent}.

From studying the mirror-field entanglement as a function of the various parameters of the model pertaining to the \textit{idf} and otherwise, we find that the presence of the \textit{idf} can influence the entanglement dynamics to a significant extent. Not only does a microscopic model like  the MOF model reproduce the known optomechanical properties and provide a more self-consistent approach of studying optomechanical interactions, more importantly it  leads to qualitatively different physics, specifically in the quantum regime. We conclude that the internal degree of freedom being the quintessential mediator of quantum correlations between the mirror center of mass and the field, the MOF model gives a physically more complete treatment of the mirror-field entanglement.

\section {Acknowledgements}
KS would like to thank Yigit Subasi and  Nick Cummings for insightful discussions and technical help through the course of the work and helpful correspondences with Ryan Behunin and Chad Galley on specific points. She acknowledges support from the Physics Frontier Center at the JQI for part of the research. SYL was supported by the Ministry of Science and Technology of Taiwan under grant MOST 102-2112-M-018-005-MY3, MOST 103-2918-I-018-004, and in part by the National Center for Theoretical Sciences, Taiwan. Part of this work was carried out when BLH was visiting the Center for Theoretical Physics of Fudan University, Shanghai, China in the summer of 2014. 

\appendix
\section{Logarithmic Negativity}
\label{LN}
After $t=0$ the interaction is turned on and the three subsystems (\mdf, \textit{idf} and the field) begin to interact with each other as the reduced density matrices for each of individual becomes a mixed state. The linearity of the interaction terms guarantees that the quantum state of the three harmonic oscillators that starts Gaussian remains Gaussian. Thus the dynamics of quantum entanglement can be studied by examining the behavior of the quantity $\Sigma$ \cite{LCH08} and the logarithmic negativity $E_{MF}$ \cite{VW02}:
\begin{eqnarray}
  \Sigma &\equiv&\det\left[ {\mathcal V_{MF}}^{PT}+{i\hbar\over 2}{\mathcal M}\right],\\
  E_{MF} &\equiv& \max \left\{ 0, -\log_2 2c_- \right\}.
\end{eqnarray}

Here ${\mathcal M}$ is the symplectic matrix ${\bf 1}\otimes (-i)\sigma_y$, $\mathcal V_{MF}$ is the partial transpose of the covariance matrix
\begin{align}
\mathcal V_{MF}= \left(\begin{array}{cc}
\bf {V}_M &\bf V_{MF} \\
\bf {V}^T_{MF} & \bf V_{F}
\end{array}\right)
\end{align}
as defined in \eqref{cmf}. $(c_+, c_-)$ is the symplectic spectrum of ${\mathcal V_{MF}}^{PT}+ (i\hbar/2){\mathcal M}$, given by
\begin{equation}
  c_\pm \equiv \left[Z \pm \sqrt{Z^2-4\det {\mathcal V_{MF}}}\over 2
    \right]^{1/2}
\label{SympSpec}
\end{equation}
with
\begin{equation}
  Z = \det {\bf V}_{MM} + \det {\bf V}_{FF} - 2 \det {\bf V}_{MF}.
\end{equation}

For the quantum oscillators in Gaussian state, $E_{MF}>0$, $\Sigma <0$, and $c_- < \hbar/2$, if and only if the quantum state of the two subsystems is entangled \cite{Si00}. $E_{MF}$ is an entanglement monotone \cite{Plenio05} whose value can indicate the degree of entanglement.

\section{Comparison of the MOF model with boundary conditions}
\label{compareBC}
Let us start with the standard optomechanical treatment where we treat the mirror field interaction via the radiation pressure. In 1+1 D, the Hamiltonian of a scalar field interacting (that corresponds to the vector potential of the optical field) interacting with a point-like mirror is given by
\begin{align}
\tilde H_{BC} = \frac{\hbar \mho}{2}\bkt{\boldsymbol P^2+\boldsymbol Z^2}+\int_0^L\dd x\bkt{\frac{\tilde\Pi^2}{2\epsilon_0}+\frac{\epsilon_0 }{2}c^2\bkt{\partial_x\tilde\Phi}^2 -\frac{\epsilon_0c^2}{2}\partial_x\bar\Phi\partial_x \tilde\Phi Z_{ZPM}\boldsymbol Z\delta(x-\bar Z)}
\end{align}
Again, assuming that a single field mode is being driven at a frequency $\omega$, we have
\begin{align}
\tilde H_{BC}=\frac{\hbar \mho}{2}\bkt{\boldsymbol P^2+\boldsymbol Z^2}+\hbar\omega \boldsymbol a^\dagger \boldsymbol a -\hbar\bkt{\frac{ \omega}{L}Z_{ZPM}A_0}(\boldsymbol a e^{i\phi_0}+\boldsymbol a^\dagger e^{-i\phi_0})\boldsymbol Z
\end{align}
The pre-factor $\beta_{MF}\equiv\frac{\omega}{L}Z_{ZPM}A_0e^{i\phi_0}$ in the interaction term is the standard optomechanical coupling as in  \cite{Vitali07} with $Z_{ZPM}\equiv \sqrt{\frac{\hbar}{M\mho}}$ as the zero point motion length for the center of mass motion and $A_0$ as the dimensionless field amplitude. Moving to a rotating frame with respect to the free field Hamiltonian $\bkt{\boldsymbol H_F \equiv \hbar \omega \boldsymbol a^\dagger\boldsymbol a}$ leads us to the following equations of motion for the dimensionless field and mirror variables
\begin{align}
\der{\boldsymbol Z}{t} &= \mho\boldsymbol P \\
\der{\boldsymbol P}{t} &= -\mho\boldsymbol Z+\sqrt 2\re\beta_{MF}\boldsymbol \Phi-\sqrt 2\im\beta_{MF}\boldsymbol \Pi\\
\der{\boldsymbol \Phi}{t} &=\sqrt 2 \im\beta_{MF}\boldsymbol Z\\
\der{\boldsymbol \Pi}{t} &= \sqrt 2\re \beta_{MF}\boldsymbol Z
\end{align}
It can be seen that in the weak coupling limit $\bkt{\lambda^2\ll2m\omega\epsilon_0c}$, redefining the field amplitude in terms of the dimensionless amplitude $A_0$ as $\Phi_0 = \sqrt{\frac{\hbar}{2\omega \epsilon_0 L}}A_0$, the effective mirror-field coupling coefficient  in \eqref{alphamf} reduces to $\alpha_{MF}\approx-\beta_{MF}/\sqrt 2 $. Solving for the entanglement in the two cases, including a damping coefficient for the internal degree of freedom and setting the {\it idf} detuning to be large we see a perfect overlap of the log negativity from either approach in Fig.\ref{comp}.

\vspace{-0.2cm}\footnotesize{{

}
\end{document}